\documentclass[10pt,letterpaper]{article}
\usepackage[top=0.85in,left=2.75in,footskip=0.75in]{geometry}

\usepackage{amsmath,amssymb}
\usepackage{setspace}

\usepackage{changepage}

\usepackage{textcomp,marvosym}

\usepackage{cite}
\usepackage{bbm}

\usepackage{nameref,hyperref}

\usepackage[right]{lineno}

\usepackage[nopatch=eqnum]{microtype}
\DisableLigatures[f]{encoding = *, family = * }

\usepackage[table]{xcolor}

\usepackage{array}

\newcolumntype{+}{!{\vrule width 2pt}}

\newlength\savedwidth

\raggedright
\usepackage[aboveskip=1pt,labelfont=bf,labelsep=period,justification=raggedright,singlelinecheck=off]{caption}

\makeatletter
\renewcommand{\@biblabel}[1]{\quad#1.}
\makeatother

\usepackage{lastpage,fancyhdr,graphicx}
\usepackage{epstopdf}
\fancyheadoffset[L]{2.25in}
\fancyfootoffset[L]{2.25in}
\begin{document}
\vspace*{0.2in}

\begin{flushleft}
{\Large
\textbf\newline{FastEnsemble: scalable ensemble clustering on large networks} 
}
\newline
\\
Yasamin Tabatabaee\textsuperscript{1},
Eleanor Wedell\textsuperscript{1},
Minhyuk Park\textsuperscript{1},
Tandy Warnow\textsuperscript{1*}
\\
\bigskip
\textbf{1} Siebel School of Computing and Data Science, University of Illinois Urbana-Champaign, Urbana, Illinois, United States of America
\\
\bigskip

* warnow@illinois.edu

\end{flushleft}
\section*{Abstract}
Many community detection algorithms are inherently stochastic, leading to variations in their output depending on input parameters and random seeds.
This variability makes the results of a single run of these algorithms less reliable. Moreover, different clustering algorithms, optimization criteria (e.g., modularity, the Constant Potts model), and resolution values can result in substantially different partitions on the same network.
Consensus clustering  methods, such as ECG and FastConsensus, have been proposed to reduce the instability of non-deterministic algorithms and improve their accuracy by combining a set of partitions resulting from multiple runs of a clustering algorithm.
In this work, we introduce FastEnsemble, a new consensus clustering method.
Our results on a wide range of synthetic networks show that FastEnsemble produces more accurate clusterings than two other consensus clustering methods, ECG and FastConsensus, for many model conditions.
Furthermore, FastEnsemble is fast enough to be used on networks with more than 3 million nodes, and so improves on the speed and scalability of FastConsensus. Finally, we showcase the utility of consensus clustering methods in mitigating the effect of resolution limit and clustering networks that are only partially covered by communities.

\section*{Author summary}
Consensus (ensemble) clustering methods, such as FastConsensus and ECG, combine partitions from multiple runs of the same clustering algorithm, in order to improve stability and accuracy of the output partition. In this study, we present a new ensemble clustering method, FastEnsemble, and show that it provides improved accuracy under many conditions compared to FastConsensus and ECG.
We show results using FastEnsemble with Leiden optimizing modularity or the Constant Potts model (CPM) and the Louvain   algorithm on synthetic networks. We show that FastEnsemble and other consensus clustering methods can reduce the effect of resolution limit for both modularity- and CPM-optimization.
Finally, we demonstrate that consensus clustering methods can improve  community detection over modularity-optimization using Leiden  on networks with both clusterable and unclusterable regions.


\section*{Introduction}
Community detection methods are commonly used to analyze the community structure of complex networks, where a community is a set of nodes that satisfies criteria such as being  dense (more edges than expected),   well-connected (i.e., not having a small edge cut) \cite{park2024well-journal}, and reasonably separable from the rest of the network.
Over the past few decades, numerous community detection methods have been developed \cite{yang2016comparative,fortunato2010community}, most of which rely on heuristic techniques  for NP-hard optimization problems, such as modularity optimization \cite{newman2004finding} or optimization under the Constant Potts model (CPM) \cite{ronhovde2010local}.

One difficulty in using community detection algorithms is that most do not produce a unique output on the same network in multiple runs.  In most cases, this variability arises from the stochastic nature of the algorithm, which incorporates randomness in the clustering process. As a result, the output can vary depending on factors such as random seeds, initial conditions, and tie-breaking rules used in the algorithm \cite{lancichinetti2012consensus,morea2024enhancing}. For instance, in the Leiden algorithm \cite{traag2019louvain}, the ordering of nodes and changes in random seeds can substantially affect the final clustering  \cite{boyack2022improved}.

On the other hand, even when the clustering algorithm is deterministic, the output may vary based on the specific optimization criterion being used (e.g. modularity or CPM) or the scale at which the clustering is done (i.e. resolution parameter).  In many cases, it is not immediately clear which optimization function or algorithmic parameters—such as the resolution parameter in Leiden or the value of $k$ in Iterative k-core Clustering (IKC) \cite{wedell2022center} will yield the best partition for a given network.  This unpredictability, combined with the challenge of selecting the most suitable optimization criteria and parameters, highlights the need for systematic methods to evaluate and compare partitions, either qualitatively or quantitatively. Alternatively, combining information from multiple partitions can lead to a more robust and representative community structure of the network.

To address these challenges, consensus (or ensemble) clustering approaches have been proposed \cite{strehl2002cluster,lancichinetti2012consensus,tandon2019fast,jeub2018multiresolution,goder2008consensus,li2008weighted,lock2013bayesian, van2022fast,zhang2014scalable,hussain2025parallel} with the goal of reducing the noise in the final clustering, which arises from the stochasticity of methods. Previous studies have shown that these consensus approaches could lead to more robust and stable partitions, and improve the accuracy of the output clustering \cite{tandon2019fast,lancichinetti2012consensus,jeub2018multiresolution}.

A class of consensus clustering methods, introduced in \cite{lancichinetti2012consensus}, take a network $G$ as input and run a clustering algorithm (such as Louvain \cite{de2011generalized,que2015scalable-louvain} with different random seeds) on it $n_p$ times to get $n_p$ different partitions. These partitions are then analyzed to construct a co-classification matrix, which captures how frequently each pair of nodes are co-clustered.  Using this matrix, a new weighted network $G^\prime$ is created and subsequently re-clustered $n_p$ times. This iterative process continues until $G^\prime$ stabilizes, converging to a stationary network. Several variations of this consensus approach have been proposed in the literature \cite{tandon2019fast, lancichinetti2012consensus, jeub2018multiresolution}.

Scalability remains a challenge for these approaches, as constructing the co-classification matrix is computationally intensive when the network is large. FastConsensus \cite{tandon2019fast} addresses this issue by employing a sampling technique, computing the co-classification matrix only for a subset of node pairs. Another recent and promising consensus method is Ensemble Clustering for Graphs (ECG) \cite{poulin2019ensemble}, which simplifies the process compared to FastConsensus by combining partitions in a single step rather than through iterative refinements.

In a recent paper published in Complex Networks and Their Applications 2024 \cite{tabatabaee2024fastensemble}, we introduced FastEnsemble, a new ensemble clustering method.
While FastEnsemble shares design similarities with ECG and FastConsensus,
FastEnsemble is designed to support both modularity and CPM optimization, whereas ECG and FastConsensus only work with modularity optimization.
Additionally, FastEnsemble eliminates much of the technical complexity of FastConsensus, allowing it to scale efficiently to large networks. To evaluate its performance, we tested a simplified version of FastEnsemble using Leiden with both modularity and CPM-based optimization on large synthetic networks generated with the LFR benchmark software \cite{lancichinetti2008benchmark,lfr-generation-code}. We compared FastEnsemble against FastConsensus and ECG in terms of accuracy and scalability and demonstrated cases where FastEnsemble provided an advantage over these methods.. Furthermore, we demonstrated that consensus clustering methods can help mitigate the resolution limit problem \cite{fortunato2007resolution} and improve clustering accuracy on networks where only a portion of the network has community structure.

In this extended study, we expand on our previous work in several directions. While \cite{tabatabaee2024fastensemble} focused on a limited set of networks for algorithm design experiments, varying only in terms of the mixing parameter, we expand our analysis to include synthetic networks with a wider range of densities and sizes. Additionally, whereas the original study demonstrated the impact of the resolution limit only for modularity-based optimization, we show that CPM-based optimization is also susceptible to the resolution limit at sufficiently small resolution values. We further extend our experiments by incorporating additional networks, including those partially composed of Erdős-Rényi graphs and tree-of-cliques. Finally, unlike \cite{tabatabaee2024fastensemble}, which  only used Leiden for modularity optimization, we include the Louvain algorithm in our experiments to ensure a fair comparison with ECG and FastConsensus, both of which also utilize Louvain.

\section*{Preliminaries}
\label{sec:prelim}

In this section, we introduce the notations and concepts related to networks and clustering used throughout this paper.

\paragraph{Notation and definitions.} Let $N = G(V, E)$ be a network where $V$ denotes  the set of nodes and $E$ denotes the  set of edges, and let $n = |V|$ and $m = |E|$.
A partition or clustering $\mathcal{P}$ of $N$ divides the set of nodes $V$ into $k$ non-overlapping sets $C_1, C_2, \dots, C_{k}$ such that each vertex belongs to exactly one cluster. We use clustering and partition interchangeably throughout this paper, and refer to each $C_i$ as a cluster or community.

In a synthetic network, we will have known ground truth communities.  In this study we will constrain all such communities to be internally connected, i.e., to not be comprised of two or more components.

For a fixed partition $\mathcal{P}$, let $d_i^{in}$ indicate the degree of node $v_i$ inside its own community and $d_i^{out}$ indicate the degree of $v_i$ outside its community. The total degree of $v_i$ is therefore $d_i = d_i^{in} + d_i^{out}$. The estimated \textit{mixing parameter} of the network for the partition $\mathcal{P}$ is defined as
\begin{equation}
    \hat{\mu}_{(N, \mathcal{P})} =  \frac{1}{n}\sum_{i \in \{1, \dots, n\}} \frac{d_i^{out}}{d_i^{in} + d_i^{out}},
\label{eq:mp}
\end{equation}
\noindent
which is equivalent to the average ratio of the number of neighbors of a node outside its community to its total degree.
When $\mathcal{P}$ is the ground-truth community structure of a network, the mixing parameter serves as an indicator of clustering difficulty for that network;  small mixing parameters signify networks that are generally easy to cluster \cite{lancichinetti2008benchmark}, whereas large mixing parameters correspond to  networks that have less clear boundaries around their clusters and are therefore more difficult to cluster correctly.

For a set of partitions $\mathcal{P}_1, \mathcal{P}_2, \dots, \mathcal{P}_{n_p}$ on network $N$ with $n$ vertices, the \textit{co-classification} or \textit{consensus} matrix $A$ is an $n\times n$ matrix where each row and each column corresponds to a vertex in $N$. The entry $A_{ij}$ represents the proportion of the $n_p$ partitions in which nodes $v_i$ and $v_j$ are co-clustered together \cite{lancichinetti2012consensus}.

\section*{Fast Ensemble Clustering}
\label{sec:method}

In this section, we describe the algorithm and implementation of FastEnsemble.

\paragraph{Algorithm.} In its simplest form, FastEnsemble uses three main parameters: an integer $n_p$ that indicates the number of partitions, a threshold $0 \le t \le 1$ for removing weak edges in the consensus matrix, and the clustering method. Given an input network $N$, FastEnsemble uses  the specified clustering method to generate $n_p$ partitions of  $N$, and creates a co-classification matrix based on these partitions.
 It then builds a new network on the same node and edge set but with the edges weighted by the entries in the co-classification matrix, i.e., the fraction of the clusterings in which the endpoints of an edge are in the same cluster. If a given edge has weight less than $t$, then the edge is removed from the network; hence the new network can have fewer edges than the original network.
The new  network is then clustered once more,  using the selected clustering algorithm, and with the option of using the weights on the edges.
Increasing the number $n_p$ of partitions can enhance accuracy and stability but comes with a computational cost. To ensure scalability for large networks, we set the default value of $n_p =10$.
We choose the default value for the parameter $t$ based on a set of algorithm design experiments.
The runtime of  FastEnsemble is $O(n^2)$.

In its advanced mode,  FastEnsemble can integrate the outputs of different clustering algorithms with arbitrary weights. 
Thus, FastEnsemble can take as input a set of clustering methods $M_1, M_2, \dots, M_{n_p}$, a set of weights $w_1, w_2, \dots, w_{n_p}$ and a set of parameters $r_1, r_2, \dots, r_{n_p}$ as input, in addition to the threshold $t$. The parameter $r_i$ represents a relevant setting or parameter for the clustering method $M_i$. For example, $r_i$ can be the resolution parameter when using the Leiden  algorithm for CPM-optimization. Finally, $w_i$ defines the weight of the clustering method $M_i$, representing its relative influence on the final clustering output. The entries of the co-classification matrix are adjusted based on these weights, so that
\begin{equation}
    A_{ij} = \frac{\sum_{t \in \{1, \dots, n_p\}} w_t
 \mathbbm{1}|v_i \text{ and } v_j \text{ are co-clustered in }C_t|}{n_p\sum w_t }
\end{equation}

\paragraph{Strict Consensus.}
We refer to a special case of FastEnsemble that uses $t=1$ as {\em Strict Consensus Clustering} in the experiments. In this variant, an edge remains in the weighted network only if it appears in all $n_p$ partitions, and hence all values below 1 in the co-classification matrix will be removed.

\paragraph{Implementation.}
FastEnsemble is a generalized framework that can be used with one or a combination of clustering paradigms with customizable weights. It is currently implemented for use with Leiden optimizing modularity (referred to as ``Leiden-mod" in the experiments), Leiden optimizing CPM (``Leiden-CPM"), and the Louvain algorithm, which optimizes modularity. However, additional clustering methods can be easily incorporated into the implementation.

\section*{Performance Study}
\label{sec:study}

\subsection*{Networks}
\label{sec:networks}

We used a selected set of synthetic networks, some available from prior studies, and some generated for this study.
Table \ref{table:datasets} provides a summary of empirical statistics of the synthetic networks, including network size and mixing parameters \cite{newman2003mixing} (see also Fig A in S1 Appendix).
Networks that have  mixing parameters of $0.5$ or larger are considered challenging to cluster
while networks with much smaller mixing parameters are generally easy to cluster  \cite{lancichinetti2008benchmark,jiang2020community}.

\paragraph{Algorithm design experiments.}

For the algorithm design experiment, we generated LFR networks using parameters similar to those used in \cite{tandon2019fast}, but with a modified exponent for the cluster size distribution to better match properties of real-world networks (see also Sec A.1.1  and Fig A in S1 Appendix).
The default model condition in this dataset consists of synthetic networks with 10,000 nodes, an average degree of 10, and estimated mixing parameter values ranging from $0.196$ to $0.978$ (note that the model mixing parameters, which are used to generate the networks, are drawn from $0.1, 0.2, \ldots, 0.9$, but the resultant mixing parameters are different).
We vary the network density (i.e., average node degree) between 5 and 20 and number of nodes between 1,000 and 100,000 to create additional networks that allow us to evaluate FastEnsemble under a range of model conditions.
In total, the algorithm design dataset has 45 model conditions, and each model condition has one replicate (i.e., one network).
These networks have mixing parameters between 0.195 to 0.978 (Table \ref{table:datasets}).

\paragraph{Testing experiments.}
The testing experiments used several different sets of synthetic networks.
One set, taken from  \cite{park2024well-journal}, contains LFR \cite{lancichinetti2008benchmark} networks based on parameters obtained from five  real-world networks clustered using Leiden-mod or Leiden-CPM.
The  five real-world networks are cit\_hepph, the Curated Exosome Network (CEN), Open Citations (OC), wiki\_topcats, and cit\_patents.
Two of these LFR networks based on CPM clustering  contained a large percentage of ground truth clusters that were internally disconnected and were therefore excluded from the experiments in \cite{park2024well-journal} as well as from this study. Additionally, LFR failed to generate a network for one model condition from a Leiden-CPM clustering.
Thus, there are 5 LFR networks that are based on Leiden-mod clusterings and 22 LFR networks that are based on Leiden-CPM clusterings.
The networks based on Leiden-mod clusterings have small mixing parameters ranging from $0.114-0.199$, while the networks based on Leiden-CPM clusterings have mixing parameters that range from $0.086$ to $0.871$ (Table \ref{table:datasets}).
The five LFR networks based on Leiden-mod clusterings of real-world networks are used to evaluate the modularity-based consensus clustering methods ECG, FastConsensus, and FastEnsemble using Leiden-mod in Experiment 2.
The 22 networks based on Leiden-CPM clusterings are used to evaluate FastEnsemble using Leiden-CPM, in comparison to Leiden-CPM, in Experiment 3.

We also included  LFR synthetic networks where the resolution limit \cite{fortunato2007resolution} is known to cause a problem for modularity-based clustering; these were used to evaluate both modularity-based clusterings and CPM-based clusterings in Experiment 4.
There are 6 ring-of-cliques networks and 5 tree-of-cliques networks in this collection.
The ring-of-cliques networks have $n$ cliques of size $10$, each connected to the cliques on the two sides by a single edge.
The tree-of-cliques networks are formed by taking a random tree on $n$ nodes and replacing each node by a clique of size $10$.
The mixing parameters for these networks are very small, in the $0.018-0.02$ range (Table \ref{table:datasets}).

We studied 14 networks that have at least half of the nodes not in any clusters in Experiment 5.
Some of these networks are Erd\H{o}s-R\'enyi graphs \cite{erdos-renyi}, and others are hybrid
 networks that contain Erd\H{o}s-R\'enyi graphs as subnetworks.
By construction, half of each hybrid network  has no community structure (i.e., every node is in a singleton cluster) and the other half has very strong community structure, as reflected by a very low mixing parameter.
 The combination of these two subgraphs produces
 mixing parameters  in the range $0.40-0.572$ (Table \ref{table:datasets}).



\begin{table}
    \caption{\textbf{Empirical statistics of the synthetic networks used in this study.}
   }
\begin{center}
\begin{small}
\begin{tabular}{||c | c c c c c ||}
 \hline
 Network & Expt. & nodes & edges & mixing param. & publ.\\ [0.5ex]
 \hline
LFR algorithm design & 1 & 1,000-1,000,000 & 5708-600227  & 0.195-0.978  & this study \\
LFR cit\_hepph MOD & 2 & 34,546 & $\sim 431K$ & 0.155 & \cite{park2024well-journal}\\
LFR wiki\_topcats MOD &2 & 1,791,489 & $\sim 24M$ & 0.199 & \cite{park2024well-journal}\\
LFR cen MOD & 2 & 3,000,000 & $\sim 21M$ & 0.180 & \cite{park2024well-journal}\\
LFR OC MOD & 2 & 3,000,000 & $\sim 55M$ & 0.129 & \cite{park2024well-journal}\\
LFR cit\_patents MOD &2&  3,774,768 & $\sim 16M$ & 0.114 & \cite{park2024well-journal}\\
LFR cit\_hepph CPM & 3 & 34,546 & $\sim 431K$ & 0.086-0.781 & \cite{park2024well-journal}\\
LFR wiki\_topcats CPM &3 & 1,791,489 & $\sim 24M$ & 0.379-0.793 & \cite{park2024well-journal}\\
LFR cen CPM & 3 & 3,000,000 & $\sim 21M$ & 0.402-0.646 & \cite{park2024well-journal}\\
LFR OC CPM & 3 & 3,000,000 & $\sim 55M$ & 0.407-0.871 & \cite{park2024well-journal}\\
LFR cit\_patents CPM &3&  3,774,768 & $\sim 16M$ & 0.211-0.807 & \cite{park2024well-journal}\\
Ring-of-cliques & 4& 90-10,000 & 4140-460,000 & 0.02 & this study \\
Tree-of-cliques & 4& 90-5,000 & 4139-229,999 & 0.018 & this study \\
Erd\H{o}s-R\'enyi & 5 & 1000 & 470-50,025 &  0.625-1.0 & this study\\
Erd\H{o}s-R\'enyi+LFR & 5 &  2000 & 4776-53,917 & 0.486-0.572 & this study\\
Erd\H{o}s-R\'enyi+ring & 5 & 2000  & 5100-54,470  & 0.40-0.51 & this study\\
 [1ex]
 \hline
\end{tabular}
\end{small}
\end{center}
\label{table:datasets}
\begin{minipage}{13cm}
\vspace{0.1cm}
\small  Notes:
 We report the number of nodes, number of edges, and the range of mixing parameters for each network collection. The mixing parameter is measured for the ``ground truth" community structure; for Erd\H{o}s-R\'enyi  graphs, we assume the ground truth clustering has each node forming its own community.
    The  rows for Experiment 3 each represent up to five different networks, each generated based on a Leiden-CPM clustering with different resolution values of the specified real-world network.
    \end{minipage}
\end{table}
\subsection*{Methods}

We evaluate FastEnsemble, ECG, and FastConsensus used with base methods for modularity optimization (Louvain for FastEnsemble and ECG, and Leiden-mod for FastConsensus).
Since ECG and FastConsensus consider weighted edges, for these analyses we also run FastEnsemble with weights on the edges in the final clustering step.
We include Leiden-mod for a baseline comparison.
We also evaluate FastEnsemble used with Leiden-CPM as the base method, and compare it to Leiden-CPM;   to keep this comparison
simple, we do not include weights on the edges in the final clustering step for FastEnsemble.

ECG is  similar to FastEnsemble, as both follow a two-step process: first, generating multiple partitions using a clustering algorithm with different random seeds, and second, combining these partitions into a final clustering by assessing the fraction of times each node pair is co-clustered and then applying a clustering algorithm to the resulting weighted network. However, there are two key differences between ECG and FastEnsemble.

First, ECG assigns a predefined minimum edge weight to edges that are not part of a 2-core (i.e., a subnetwork where every node is adjacent to at least two other nodes in the subnetwork) in the original graph. In contrast, FastEnsemble assigns high weights to edges whose endpoints are frequently co-clustered across partitions, regardless of their inclusion in a 2-core. As a result, ECG is less likely to co-cluster node pairs in the final consensus clustering if they do not belong to 2-cores in the input network, whereas FastEnsemble does not impose this structural constraint.
Second, ECG is restricted to the Louvain algorithm (which optimizes modularity), whereas FastEnsemble has been implemented to work with Leiden or Louvain for modularity optimization, and with Leiden optimizing CPM.
Furthermore, although not examined in this study, FastEnsemble can integrate partitions from two or more clustering algorithms.

\subsection*{Evaluation criteria}

We evaluate accuracy on networks with known ground truth community structure using Normalized Mutual Information (NMI) and Adjusted Rand Index (ARI), as implemented in the Scikit-learn \cite{pedregosa2011scikit} library. Additionally, in some experiments, we report cluster size distributions to gain deeper insights into clusters.   To further assess clustering performance, we compute false negative and false positive error rates. Treating both the true and estimated clusterings as equivalence relations---each defined by a set of node pairs where
$(x,y)$ belongs to the relation if and only if nodes $x$ and $y$ are in the same cluster---we define:
\begin{itemize}
    \item False negatives (FN): Pairs present in the true clustering but missing in the estimated clustering.
    \item False positives (FP): Pairs present in the estimated clustering but absent in the true clustering.
    \item True positives (TP): Pairs present in both the true and estimated clusterings.
    \item True negatives (TN): Pairs absent from both clusterings.
\end{itemize}

Using these definitions, we report the False Negative Rate (FNR), False Positive Rate (FPR), and the F1-score computed as
\begin{equation}
    FNR = \frac{FN}{FN+TP}, \quad FPR=\frac{FP}{FP+TN}, \quad F_1 = \frac{2TP}{2TP+FP+FN}
\end{equation}

\subsection*{Experiments}

We conduct five experiments, as outlined below. In each case, we use synthetic networks and evaluate accuracy by comparing the results to the ground truth community structure. For all experiments except ones on large networks from \cite{park2024well-journal}, all analyses were allocated four hours of runtime and 64GB of memory without parallelism on the University of Illinois Campus Cluster. Any instances where a method failed to complete within this time limit were recorded.

\begin{itemize}
    \item Experiment 1: We set the default for the threshold parameter $t$ in FastEnsemble based on experiments using modularity optimization on a collection of algorithm design datasets.
    \item Experiment 2: We evaluate modularity-based consensus pipelines with respect to both accuracy and runtime on five LFR synthetic networks   from \cite{park2024well-journal}, which are  based on five real-world networks clustered using Leiden-mod. These networks have  up to $\sim 3.8$M nodes.
     \item Experiment 3: We evaluate FastConsensus used with Leiden-CPM  with respect to  accuracy and runtime on 22 LFR synthetic networks   from \cite{park2024well-journal}, which are based on  five real-world networks clustered using Leiden-CPM with different resolution parameters. These  networks have up to $\sim 3.8$M nodes.

    \item Experiment 4: We assess the robustness of different modularity-based and CPM-based clustering methods to the resolution limit using ring-of-cliques and tree-of-cliques networks with up to 100K nodes.

      \item Experiment 5: We evaluate modularity-based consensus pipelines on networks where at least half of the network  is an Erd\H{o}s-R\'enyi  graph and so the network has at most half of the nodes in non-singleton clusters.
\end{itemize}
Thus, the first four experiments examine clustering on networks where all or nearly all the nodes are in clusters of size at least two, while the last experiment examines clustering on networks where at most half of the nodes are in clusters of size at least two.
Some of these experiments   focus exclusively on modularity-based clusterings, while others examine CPM-based clusterings.
Experiments 1 and 3  use networks with a range of mixing parameters, Experiments 2 and 4 use networks with low mixing parameters, and Experiment 5 examines networks with moderate to high mixing parameters  (Table \ref{table:datasets}).

\section*{Results}
\label{sec:results}

\subsection*{Experiment 1: Algorithm design experiment}
\label{sec:expt1}

This experiment has two parts.  In Experiment 1a, we set the default value for the threshold parameter $t$ in FastEnsemble,  so that edges with support below $t$ in the co-classification matrix are removed from the weighted network.
In Experiment 1b, we compare the default setting for FastEnsemble to ECG and FastConsensus.

\subsubsection*{Experiment 1a: Setting the default threshold value}
\begin{figure}[h!]
    \centering
    \includegraphics[width=0.49\textwidth]{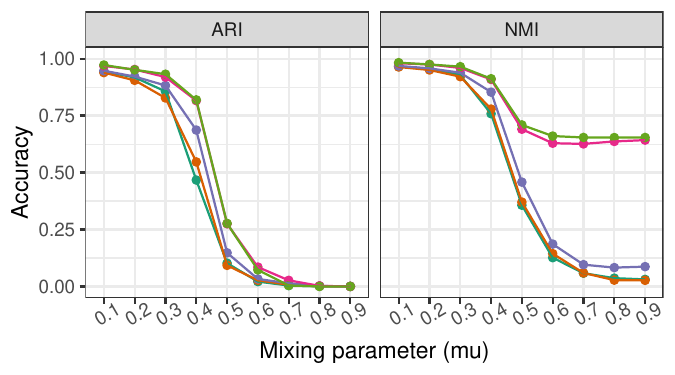}
    \includegraphics[width=0.49\textwidth]{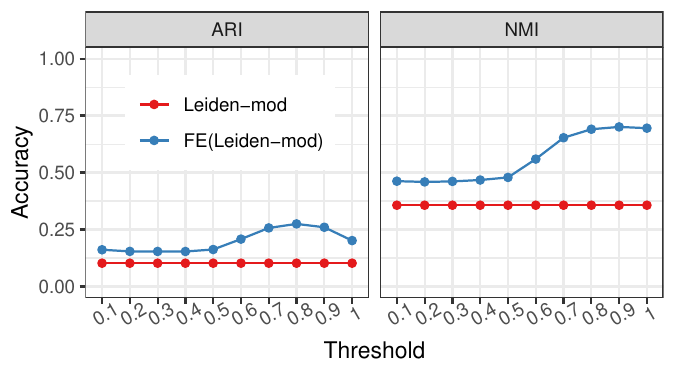}
    \includegraphics[width=0.65\textwidth]{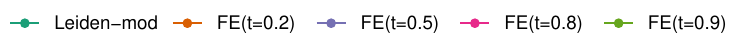}
    \caption{\textbf{Experiment 1a: Setting the default value for $t$ in  FastEnsemble.}
    Each plot shows ARI and NMI accuracy for Leiden-mod and FastEnsemble using four different threshold values on the default algorithm design networks with 10,000 nodes.
    Left: Accuracy  as a function of the model mixing parameter (x-axis).   Right: Accuracy as  a function of the threshold value on the networks   with model mixing parameter $0.5$.
    FE stands for FastEnsemble.}
    \label{fig:expt1a}
\end{figure}

As seen in Fig \ref{fig:expt1a} (left),
overall
 the best accuracy across all networks is obtained using threshold values of $t=0.8$ and $t=0.9$, with $t=0.9$ slightly outperforming $t=0.8$ in terms of NMI and  $t=0.8$ slightly outperforming $t=0.9$ in terms of ARI for moderate to high resolution values. Evaluating results on one of the LFR networks with mixing parameter of $\mu=0.5$ and allowing $t$ to vary between $0.1, 0.2, \ldots, 0.9, 1$ (Fig \ref{fig:expt1a} (right)), FastEnsemble achieves its optimal accuracy for $t$ values between $0.7$ and $0.9$ in terms of ARI and $t$  between 0.8 and 1 in terms of NMI. Overall, Fig \ref{fig:expt1a}   suggest that $t$ values of $0.8$ and $0.9$ provide the best accuracy.

 To select between these values, we evaluate variants of FastEnsemble on a wider set of conditions, varying the LFR network density (by changing the average degree), the network size as well as the mixing parameter. Increasing the network density improves accuracy for all variants when the mixing parameter is low, but results in decrease in accuracy for high mixing parameters (Fig B, S1 Appendix). 
 Increasing the network size however results in a small increase in the accuracy for the two best-performing variants (with $t=0.8$ and $t=0.9$) and a decrease (which can be large for moderate mixing parameters) for Leiden-mod and the rest of the methods (Fig B, S1 Appendix).
 In all these conditions, the two best-performing variants have a tie, but FastEnsemble with $t=0.8$  exhibits a clear advantage for mixing parameters of $0.5$ and $0.6$ for different network sizes (Fig B, S1 Appendix) and FastEnsemble with $t=0.9$ a small advantage for high mixing parameters and small network sizes. Based on these findings, we set the default threshold value for FastEnsemble to $t=0.8$, while noting that the optimal threshold may vary depending on the dataset and is influenced by the mixing parameter.

We wish to note the impact of the mixing parameter: while accuracy is very high for networks with the lowest mixing parameter, it quickly drops as the mixing parameter increases. This reflects the discussion in  \cite{jiang2020community,lancichinetti2008benchmark}.

\begin{figure}[tbp!]
    \centering
    \includegraphics[width=1\textwidth]{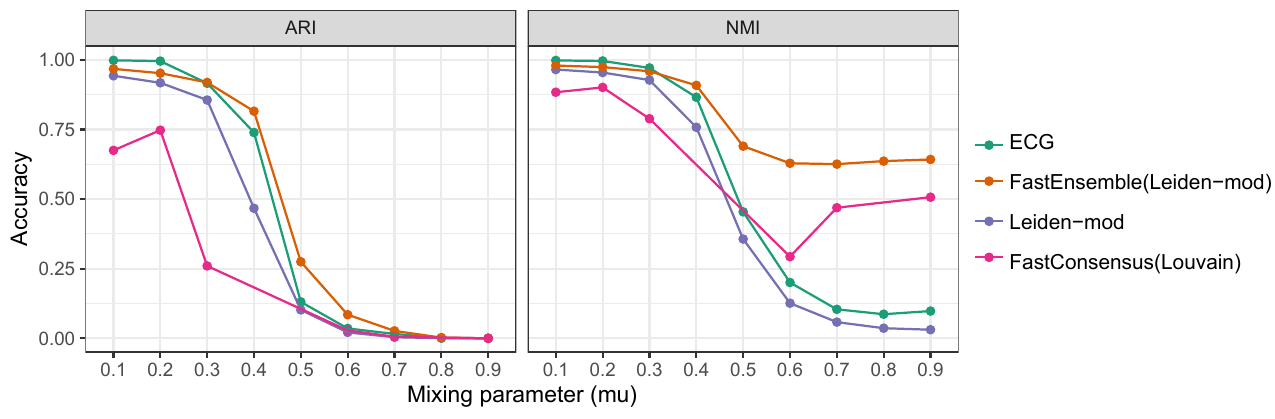}
    \caption{\textbf{Experiment 1b: Evaluating modularity-based consensus clustering pipelines on the algorithm design datasets with 10,000 nodes as a function of the mixing parameter.}  Results are shown for three consensus clustering methods and also Leiden-mod on the  algorithm design datasets with 10,000 nodes but varying mixing parameter (values on the x-axis). 
    }
    \label{fig:expt1b}
\end{figure}
\subsubsection*{Experiment 1b: Comparing default FastEnsemble to ECG and FastConsensus on Algorithm Design networks}
As seen in Fig \ref{fig:expt1b},
accuracy declines for all methods as the model mixing parameter increases.
For the default model condition, i.e., networks with 10,000 nodes and an average degree of 10, ECG achieves the highest accuracy for the two smallest mixing parameters ($0.1$ and $0.2$).
However, when the mixing parameter is  0.3 or higher, FastEnsemble outperforms the other methods. Both ECG and FastEnsemble consistently match or surpass Leiden-mod in accuracy. FastConsensus improves upon Leiden-mod for larger mixing parameters but is less accurate for smaller ones ($\mu$ values below 0.5). However, FastConsensus failed to converge in 14 out of the 45 model conditions (Table A in S1 Appendix) within the allotted four-hour time limit, including on all networks with 100K nodes.

Results across networks of varying sizes and densities exhibit the same general trend (Fig C, S1 Appendix). However, for sparse networks (average degree of 5), ECG outperforms FastEnsemble only at the smallest mixing parameter ($0.1$), while FastEnsemble is significantly more accurate for all higher mixing parameters in terms of NMI (Fig C, S1 Appendix).
On these sparse networks, ECG offers little improvement over Leiden-mod, whereas FastEnsemble achieves substantially better accuracy than both. FastConsensus's accuracy is most influenced by changes in the average degree.
For small mixing parameters, its accuracy in terms of ARI increases dramatically as network density increases. However, for large mixing values, accuracy drops sharply—when the average degree is 5, FastConsensus performs nearly identical to FastEnsemble, but at an average degree of 20, its accuracy declines substantially, reaching a level comparable to Leiden-mod.

As network size increases, ECG and Leiden-mod experience a decline in accuracy, whereas FastEnsemble improves (Fig C, S1 Appendix).
Notably, the drop in accuracy of  Leiden-mod is most pronounced for intermediate mixing parameters (0.3–0.5), which are the same conditions where FastEnsemble benefits the most from larger network sizes. These findings suggest that FastEnsemble's advantage becomes more pronounced for larger and lower-density networks. FastConsensus is less accurate than all other methods for small mixing values but outperforms ECG and Leiden-mod for larger mixing values, while consistently remaining less accurate than FastEnsemble. However, it fails to complete within four hours on all networks of the largest size (100K nodes) and under several model conditions with 10K nodes.

\subsection*{Experiment 2: Results of modularity-based clustering on LFR networks based on real-world networks }
\label{sec:expt2}

In this experiment, we evaluate the accuracy and scalability of different consensus clustering methods on the   LFR networks  from \cite{park2024well-journal} that were generated using modularity-based clusterings of five large real-world networks, which range in size from $\sim$ 35K to $\sim$ 3.8M nodes.

\begin{figure}[tbp!]
    \centering
    \includegraphics[width=1\textwidth]{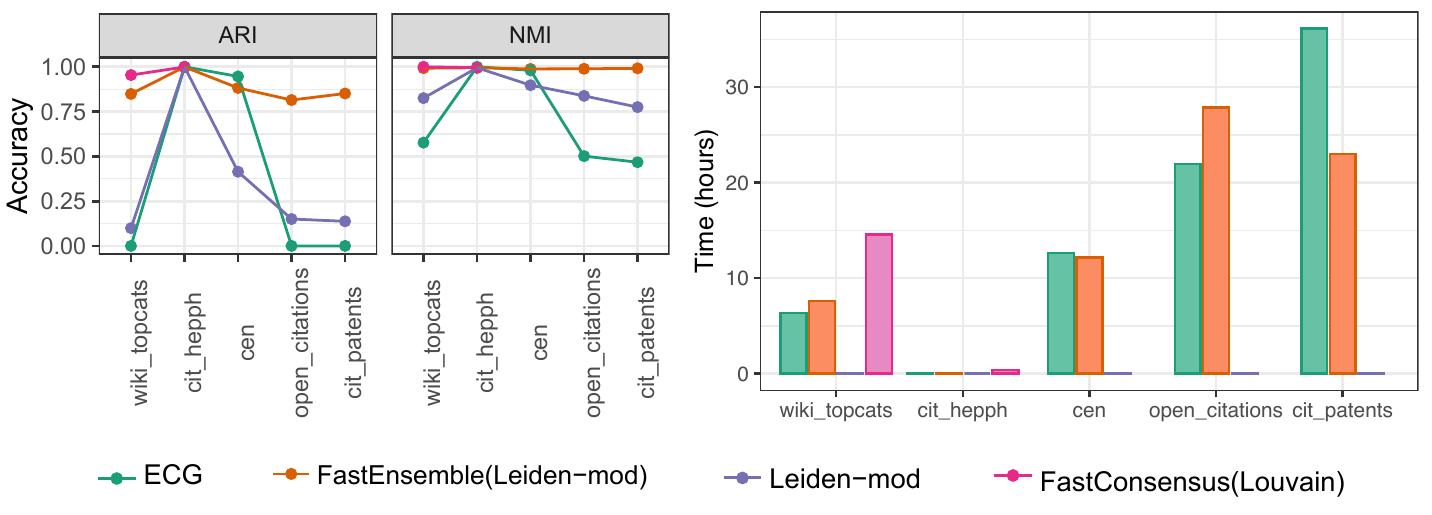}
    \caption{\textbf{Experiment 2: Evaluating modularity-based consensus clustering pipelines on synthetic networks based on clustered real-world networks.}  Results are for modularity-based clustering methods  on  LFR networks from \cite{park2024well-journal}, each  based on a Leiden-modularity clustering of a real-world network. Left: Accuracy (NMI and ARI). Right: Runtime (in hours). FastConsensus failed to converge on three networks (CEN, open\_citations, cit\_patents) within the allotted 48 hours.
    }
    \label{fig:expt2}
\end{figure}

Among all consensus clustering methods, the only network they completed on within four hours was cit\_hepph, the smallest LFR synthetic network with approximately 35K nodes. On this network, all three methods achieved near-perfect ARI and NMI scores
(Fig \ref{fig:expt2} (left), Table B in S1 Appendix).

We then extended the runtime limit to 48 hours for the four remaining networks. FastEnsemble successfully completed on all of them, taking between 7 and 28 hours. FastConsensus managed to complete on only one of these networks, requiring 14.5 hours, where it demonstrated excellent accuracy, slightly outperforming FastEnsemble. ECG completed on all networks, with runtimes ranging from 6 to 36 hours (Fig \ref{fig:expt2} (right), Table B in S1 Appendix).

Overall, FastEnsemble and ECG were both faster than FastConsensus, with FastEnsemble almost 1.5X faster on the largest network (cit\_patents). As expected, Leiden-mod was the fastest, requiring only a few seconds to minutes per network. In terms of accuracy, ECG was less accurate than FastEnsemble on three of the large networks but more accurate on one, where both methods achieved very high ARI/NMI scores, suggesting that the network was relatively easy to cluster (Table B in S1 Appendix) with an estimated mixing parameter of 0.18 \cite{park2024well-journal}.

\subsection*{Experiment 3:  Results on  CPM-based clustering on LFR networks based on clustered real-world networks}
\label{sec:expt3}

 \begin{figure}[tbp!]
    \centering
    \includegraphics[width=1\textwidth]{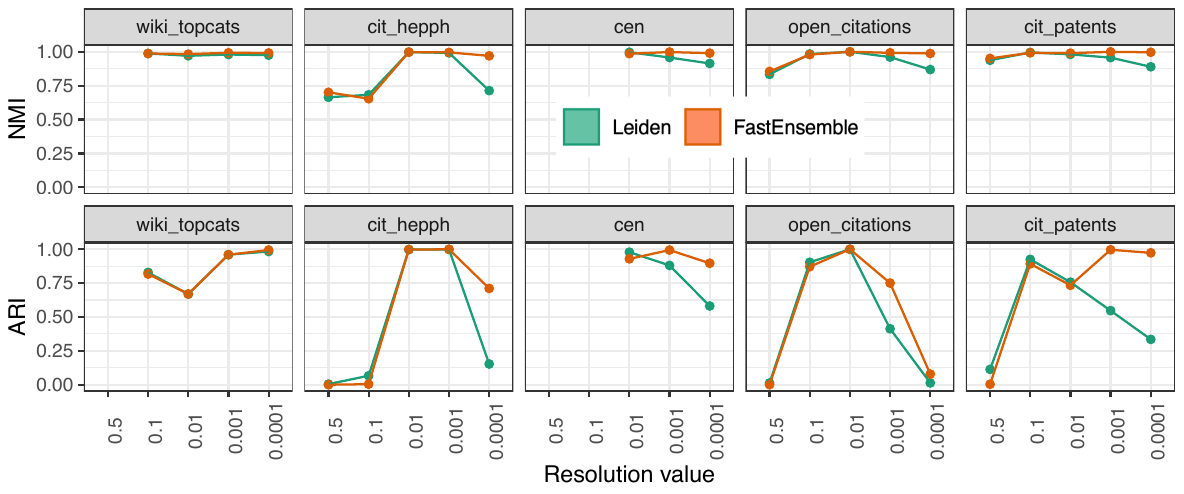}
    \includegraphics[width=0.8\textwidth]{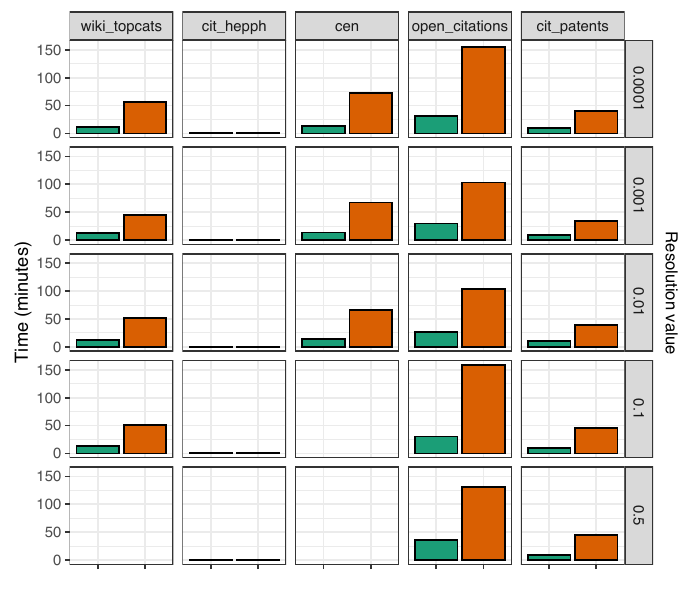}
    \caption{\textbf{Experiment 3: Comparison of  FastEnsemble(Leiden-CPM)  and Leiden-CPM  on  synthetic networks based on clustered real-world networks. }
  The LFR networks are from~\cite{park2024well-journal} and are generated from a real-world network clustered using Leiden optimizing CPM for a specific resolution parameter value.
  The clustering methods studied are Leiden-CPM and FastEnsemble using CPM, each used with the same resolution parameter value as specified for the given LFR network.
  Top: Accuracy (NMI and ARI).
  Bottom: Runtime (in minutes).
   Results are not shown for three conditions:  LFR graphs with a large fraction of disconnected ground truth clusters (the two CEN networks) or when the LFR software failed to create a network for the provided parameters (the wiki\_topcats network).}
    \label{fig:expt3-new}
 \end{figure}

In this experiment, we evaluate the accuracy and runtime of FastEnsemble using Leiden-CPM in comparison to Leiden-CPM on 22 LFR networks from  \cite{park2024well-journal}.
These networks are based on 5 real-world networks that are clustered using Leiden-CPM for varying resolution parameters, and  range in size from $\sim$ 35K to $\sim$ 3.8M nodes.
FastEnsemble consistently achieves accuracy that is at least as high as Leiden-CPM for all 22 networks and often surpasses it, particularly when used with  small resolution values (Fig \ref{fig:expt3-new}).

While Experiment 1 shows that FastEnsemble's performance gap with Leiden increases at \textit{higher} mixing parameters, in this experiment, networks generated using parameters from CPM-clustering with low resolution values correspond to \textit{lower} mixing parameters (Fig A in S1 Appendix). Despite this, FastEnsemble still achieves higher accuracy in this setting.

The comparison of runtimes shows that FastEnsemble was, on average, much slower compared to Leiden-CPM (Fig \ref{fig:expt3-new}).
Nevertheless,   FastEnsemble completes on all the networks in this collection in at most 2.5 hours (see Table C, S1 Appendix), and typically less, thus demonstrating that it can  process  large networks (up to 3.8 million nodes) in a reasonable time.

\subsection*{Experiment 4: The Resolution Limit}
\label{sec:expt4}

\begin{figure}[h!]
    \centering
    \includegraphics[width=1\textwidth]{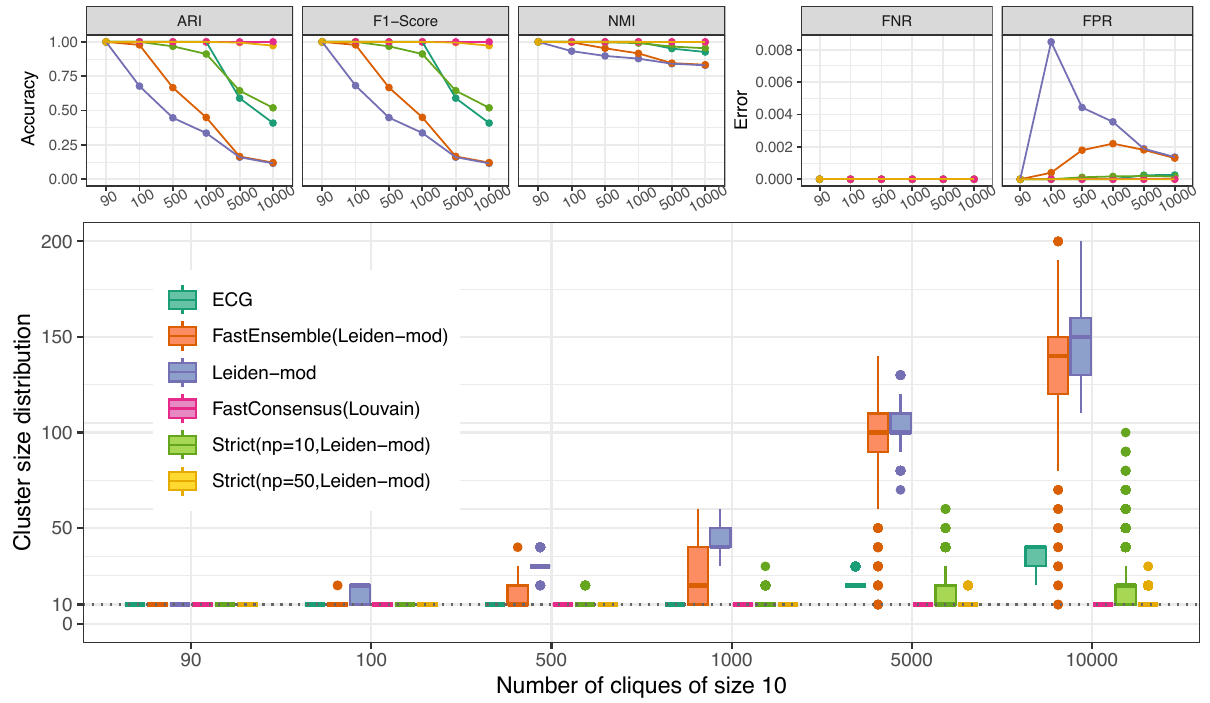}
    \caption{\textbf{Experiment 4: Accuracy of modularity-based consensus clustering methods on ring-of-cliques networks of varying sizes.}
    Each ring-of-cliques networks connects $n$ cliques of size 10 in a ring.
    The methods compared are Leiden-mod, ECG, FastConsensus, FastEnsemble, and Strict Consensus (with two numbers np of partitions).
    Top left: Accuracy (ARI, NMI, F1-score) as a function of $n$. Top right:  Error metrics (FNR and FPR) as a function of $n$. Bottom:   Cluster size distribution as a function of $n$ (the dotted line indicates the true distribution).
    }
    \label{fig:expt4-mod}
\end{figure}

The resolution limit for modularity was first defined in \cite{fortunato2007resolution}, which illustrates that in some cases, an optimal modularity-based clustering may not identify what are intuitively the ``obvious" communities, especially when those communities are small. As an example, \cite{fortunato2007resolution} proposes the family of ring-of-cliques networks, which are characterized by the clique size
$k$ and the number
$n$ of cliques. In these networks, the cliques are arranged in a ring and connected to adjacent cliques by a single edge. The study in \cite{fortunato2007resolution} shows that when
$n \geq k(k-1)+2$, the optimal modularity-based clustering will group multiple cliques into a single cluster, rather than returning the obviously preferred clustering where each clique is considered a separate community.

\begin{figure}[ht!]
    \centering
    \includegraphics[width=1\textwidth]{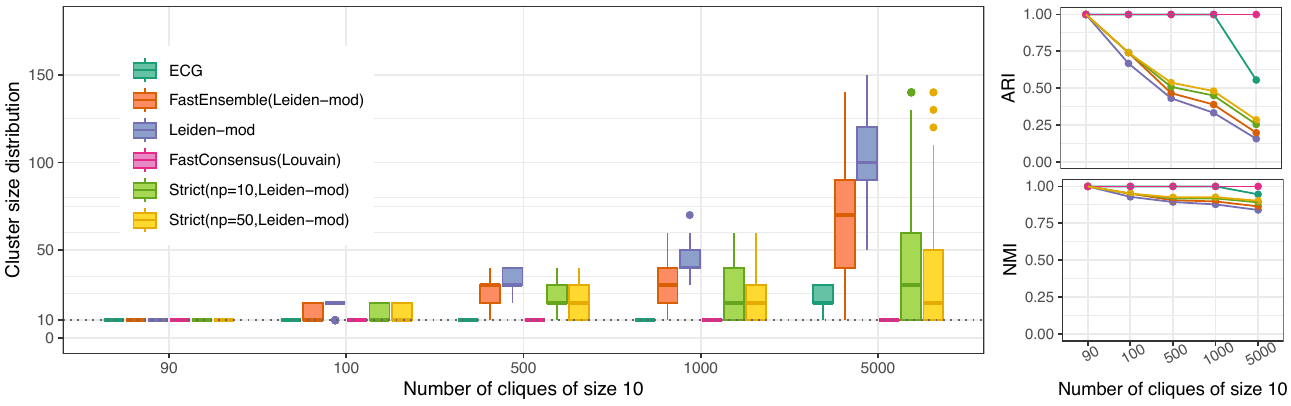}
    \caption[Impact of resolution limit on pipelines based on modularity on Tree-of-Cliques networks.]{\textbf{Experiment 4: Evaluating modularity-based  clustering methods on tree-of-cliques networks.} Each tree-of-cliques network has varying number of cliques of size 10 connected in a tree structure. Left:  Cluster size distribution, as a function of the number of cliques (dotted line is the true distribution). Right:  Clustering accuracy (ARI, NMI) as a function of the number of cliques.
    }
\label{fig:expt4-trees}
\end{figure}

\paragraph{Robustness to resolution limit for modularity optimization:}
Here we examine whether consensus clustering methods can address this vulnerability of modularity-based clustering from an empirical perspective, using  ring-of-clique networks where each clique is of size $k=10$ but the  number $n$ of cliques is allowed to vary.
According to the previous paragraph, when $n \geq 10\times 9+2=92$ then an optimal modularity clustering will group two or more of the cliques together.
Hence, we examine values of $n$ that are both smaller and larger than $n=91$ in this experiment.
The methods evaluated include Leiden-mod, FastConsensus, ECG, FastEnsemble, and two variants of Strict Consensus, differing in the number of partitions (np) used.

For $n=90$ clusters, all methods produce clusterings where each clique is returned as a separate community, as desired (Fig \ref{fig:expt4-mod}).
However, as the number of clusters increases, but not their sizes, then Leiden-mod starts merging cliques together, as predicted by the theory from \cite{fortunato2007resolution}.
We also see that the consensus clustering methods (i.e., FastConsensus, ECG, FastEnsemble, and Strict Consensus) reduce the tendency to merge cliques into clusters, but some are more beneficial than others. In particular,  the Strict Consensus variants, especially with $np=50$, have  the best accuracy, while
 FastEnsemble has poor accuracy, especially for the large numbers of clusters, where it is nearly as poor as Leiden-mod.

Note that all the methods return essentially zero FNR, indicating that no clique in the ring-of-cliques network is ever split apart (Fig \ref{fig:expt4-mod} (top right)).
On the other hand, the methods differ in terms of FPR, with Leiden-mod having high FPR  except for $n=90$.  Again, FastEnsemble  is almost as poor as Leiden-mod when there is a large number of cliques, while the other consensus methods have much lower FPR values.

Results on tree-of-cliques networks exhibit slightly different trends (Fig \ref{fig:expt4-trees}).
As with the ring-of-cliques networks, Leiden-mod has the worst accuracy, followed by FastEnsemble(Leiden-mod), and FastConsensus has the best accuracy.
However, on these networks,  ECG strictly improves on the two StrictConsensus variants, which is different from  what we saw on the ring-of-cliques networks.

\paragraph{Robustness to resolution limit for CPM-based optimization:}
\begin{figure}[h!]
    \centering
    \includegraphics[width=1\textwidth]{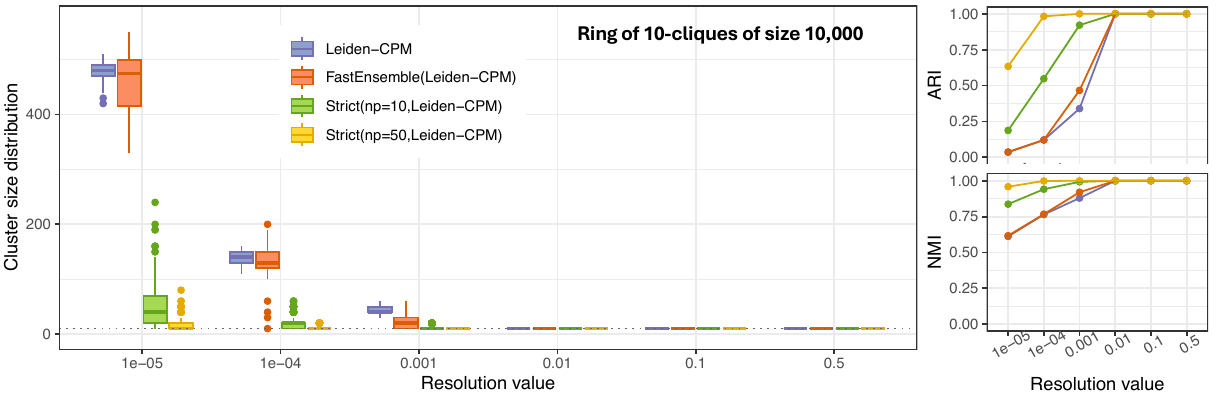}
    \includegraphics[width=1\textwidth]{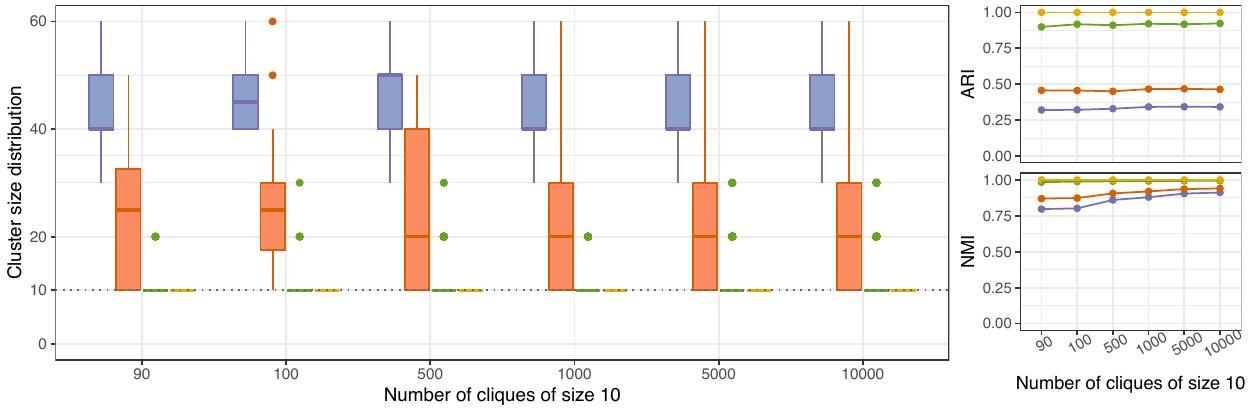}
    \caption[Experiment on ring-of-cliques networks using CPM-optimization]{\textbf{Experiment 4: Evaluating CPM-based clustering methods on ring-of-cliques networks.}
    Results are shown for Leiden-CPM, FastEnsemble with Leiden-CPM,  and the Strict Consensus with Leiden-CPM with 10 or 50 partitions (indicated by np).
    Top left: cluster size distribution as a function of the resolution parameter. Top right:  Clustering accuracy as a function of the resolution parameter. Bottom left: Cluster size distribution as a function of the number of cliques (dotted line indicates the true distribution). Bottom right: Cluster accuracy as a function of the number of cliques, using a resolution value of
$r=0.001$.
}
    \label{fig:expt4-cpm}
\end{figure}

In contrast to the theory for modularity, \cite{traag2011narrow} established that for every setting of the resolution parameter $r$, there will be a value $N$ so that every optimal CPM(r) clustering of a ring-of-cliques network with $n \geq N$ cliques of size $k$ will return the individual cliques as clusters.
However, our experimental results show that for large enough numbers of cliques of size 10 and small resolution values, Leiden-CPM groups cliques together into clusters (Fig \ref{fig:expt4-cpm}).
This vulnerability occurs for all of the small  values for the resolution parameter $r$, but disappears when $r \geq 0.01$.
Unlike in modularity-based experiments, increasing the number of cliques has little effect on the accuracy of the methods or the cluster size distribution. This suggests that, for CPM-optimization, cliques tend to be co-clustered into groups of the same size, regardless of the overall network size (e.g., Leiden-CPM returns clusters containing approximately 4 to 5 cliques when $r=0.001$, Fig \ref{fig:expt4-cpm}). Using FastEnsemble provides minimal improvement over Leiden-CPM. However, applying the Strict Consensus with Leiden-CPM resolves this issue, successfully identifying individual cliques as distinct clusters.

Note that for a ring-of-cliques network with $n$ cliques of size $k$, the mixing parameter with respect to the ground-truth community structure is equal to $\frac{2}{k^2}$ (see Sec B in S1 Appendix for derivation). When $k=10$, this value becomes $\hat{\mu}=0.02$, which agrees with Table \ref{table:datasets}. This means that the mixing parameter of a ring-of-cliques networks only depends on the size of the cliques (and not their count), and except when cliques are extremely small (e.g., at most 4 nodes), the mixing parameter is very low. The lower accuracy of FastEnsemble compared to ECG and FastConsensus in this context aligns with the findings on the algorithm design dataset.

\subsection*{Experiment 5: Clustering networks that have only partial community structure}
\label{sec:expt5}

     \begin{figure}[h!]
    \centering
     \includegraphics[width=1\textwidth]{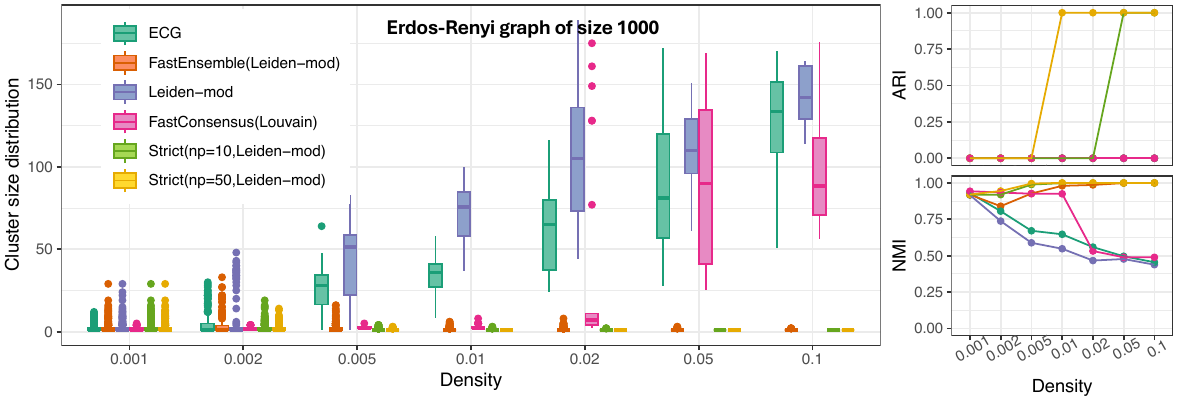}
    \includegraphics[width=1\textwidth]{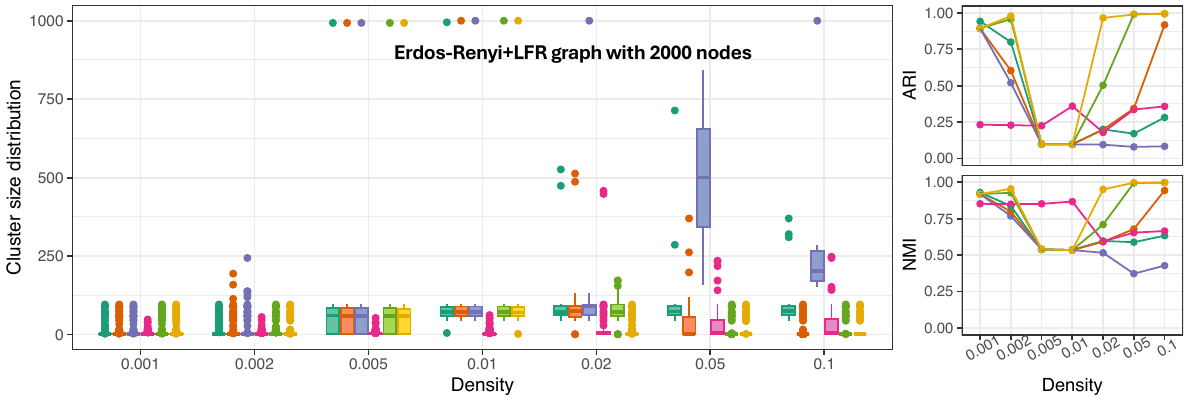}
    \includegraphics[width=1\textwidth]{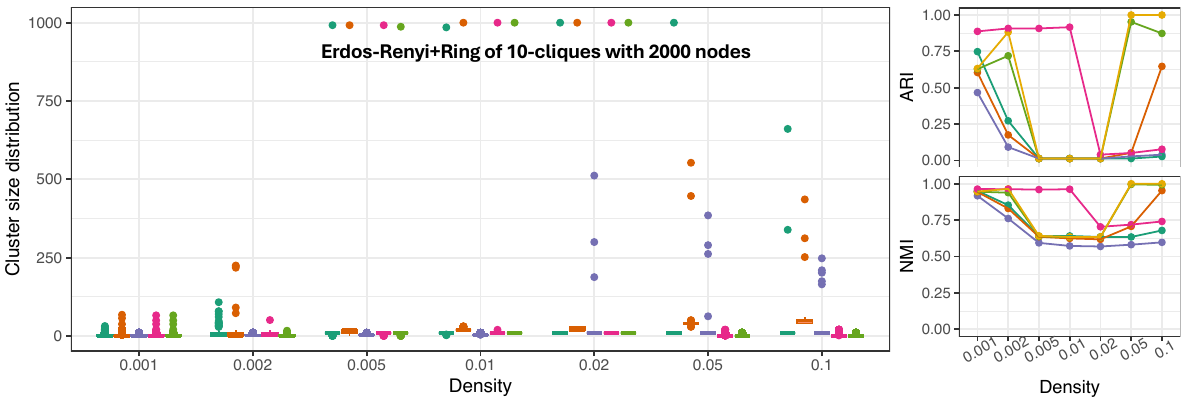}
    \caption{\textbf{Experiment 5: Clustering networks that are only partially clusterable.} Accuracy and cluster size distributions of modularity-based consensus clustering on networks with clusterable and unclusterable components. The unclusterable portion is created using Erd\H{o}s-R\'enyi  graphs with various densities, and the clusterable portion includes strong community structure created using LFR graphs or ring-of-cliques.
    Each row (top, middle, bottom) shows the cluster size distribution on the left and the clustering accuracy on the right (ARI and NMI), as a function of the density of the Erd\H{o}s-R\'enyi  graph.
    Top: Erd\H{o}s-R\'enyi  graphs with 1000 nodes and various densities. Middle:  Erd\H{o}s-R\'enyi  graph of size 1000 attached to an LFR graph of size 1000 with 14 communities (2000 nodes in total), with
    sizes ranging from 45 to 96. Bottom:  Erd\H{o}s-R\'enyi  graph attached to a ring of 10-cliques of size 1000 (2000 total nodes).
   }
\label{fig:expt5-new}
\end{figure}

While Erd\H{o}s-R\'enyi graphs may exhibit regions that appear to be valid communities based on metrics such as modularity scores, we follow the discussion in \cite{lancichinetti2011limits} and treat Erd\H{o}s-R\'enyi graphs as lacking any true community structure.
Thus, we do not consider any cluster of size greater than 1 to be valid in an Erd\H{o}s-R\'enyi graph.
We use these graphs to assess the extent to which consensus clustering pipelines can correctly reject spurious community structures by producing no or very few non-singleton clusters. Additionally, we construct hybrid networks that combine Erd\H{o}s-R\'enyi graphs with LFR networks and ring-of-cliques networks to examine whether clustering methods can correctly restrict detected communities to subnetworks with well-defined community structures. To evaluate these aspects, we analyze both the cluster size distribution and overall clustering accuracy.

On Erd\H{o}s-R\'enyi graphs, the cluster size distribution and accuracy for each method is very impacted by the density
$p$ (Fig \ref{fig:expt5-new} (top)).
In particular, while cluster sizes tend to be small for the smallest tested value for $p$,    ECG and Leiden-mod produce fairly large clusters  even at relatively small values for $p$, and  so does FastConsensus at slightly larger values.  In contrast, the clusters produced by FastEnsemble and the two StrictConsensus variants decrease in size as the density $p$ increases.
The clustering NMI and ARI accuracy results  also reflect these trends.  For ARI, all methods other than the two Strict Consensus variants have very poor accuracy at all values for $p$, but the two Strict Consensus variants improve as $p$ increases and attain high accuracy for the larger values for $p$.
NMI results show all methods have fairly high accuracy for the smallest tested value for $p$, but Leiden-Mod, ECG, and  FastConsensus degrade as $p$ increases, while FastEnsemble and the two Strict Consensus variants improve as $p$ increases.

We observe somewhat different trends in networks that combine Erd\H{o}s-R\'enyi graphs with LFR networks  (Fig \ref{fig:expt5-new} (middle)).
The LFR subnetwork has  14 ground-truth communities with sizes that range from 45 to 96 (Sec A.1.4 in S1 Appendix) and its mixing parameter is 0.14.
Therefore, the correct community structure should have half the nodes in singleton clusters and the other half in 14 clusters that do not exceed 100 nodes.
The cluster size distributions seen in Fig \ref{fig:expt5-new} (middle) seem reasonably accurate for all methods for the very lowest density values for the Erd\H{o}s-R\'enyi graphs, and then accuracy decreases.  Specifically, with the exception of FastConsensus, for the middle density values, all methods produce large clusters, and some even produce clusters of size 1000.  Upon inspection, the clusters of size 1000 were verified to be the Erd\H{o}s-R\'enyi graphs.
At the highest density values, the cluster size distribution for most methods drops closer to the true values, but Leiden-mod continues to produce very large clusters, including one of size 1000.
An examination of NMI and ARI accuracy shows interesting trends that reflect the cluster size distribution accuracy.
For ARI, the method with consistent but poor accuracy across all density values  is FastConsensus.  Leiden-mod starts with high accuracy and then drop to a very low accuracy as the density increases, and never regains accuracy. ECG is similar to Leiden-mod in starting at high accuracy and then decreasing to low accuracy, but it regains some accuracy as the density increases.
FastEnsemble and the two Strict Consensus variants show a surprising trend of starting high, dropping down to a low value, and then going back to a high value, though the Strict Consensus variants return to the high value at lower density values than FastEnsemble.
Results for NMI are similar as for ARI, but the accuracy scores are higher.

We also examined Erd\H{o}s-R\'enyi graphs combined with ring-of-cliques networks, which have mixing parameter 0.02.
On these networks, the true cluster size distribution has half the nodes in clusters of size 10 and the other half in singleton clusters.
Results on these networks are shown in  Fig \ref{fig:expt5-new} (bottom) and are similar to results on Erd\H{o}s-R\'enyi graphs with LFR networks  (Fig \ref{fig:expt5-new} (middle)).
There are  similar trends for cluster size distributions, with good accuracy at the lowest density and then all methods (other than FastConsensus)  grouping all the nodes in the Erd\H{o}s-R\'enyi graph into one cluster for the intermediate density values.  For ARI and NMI accuracy the trends are nearly identical, with one   exception: FastConsensus now has the highest ARI and NMI accuracy of all methods for the five lower density values, ties for second place on the sixth density value, and then drops to the third or fourth place for the last density value.

Overall, for this experiment the relative accuracy between methods depends very much on the density of the Erd\H{o}s-R\'enyi graph as well as the structure of the clusterable subnetwork.
 FastConsensus does poorly in one setting (when the Erd\H{o}s-R\'enyi graph is paired with LFR networks) and well in the other (when it is paired with the ring-of-cliques network).
 FastEnsemble does not have very good accuracy in either setting for middle density values, but does well at the highest density values.
 The two Strict Consensus variants, however, do well at the highest density values, and are generally more accurate than the other methods.
Nevertheless, no method does well at all density values, and no method dominates the others.

\section*{Discussion}
\label{sec:discussion}

This study evaluated the accuracy of FastEnsemble in comparison to ECG and FastConsensus  under conditions where the entire network has community structure (Experiments 1--4) or where at most half of the network has community structure (Experiment 5).

\subsection*{When the entire network has community structure}
When the entire network has community structure (Experiments 1--4),
we consistently found that FastEnsemble, ECG, and FastConsensus usually produced  clusterings that were at least as accurate as their base method (modularity- or CPM-optimization),   and sometimes were substantially more accurate
(e.g., Figs \ref{fig:expt2}--\ref{fig:expt5-new}, Table B in S1 Appendix).
The results for ECG and FastConsensus are consistent with prior studies \cite{poulin2019ECG,tandon2019fast} and are expected.

We also compared FastEnsemble to  ECG and FastConsensus.
In these experiments, which were restricted to modularity-based clustering (as ECG and FastConsensus are not designed to work with Leiden-CPM), we found cases where each had the best accuracy, so once again no method strictly dominates the other methods.
Nevertheless, we observed that the mixing parameter was a good indicator of whether FastEnsemble would be at least as accurate as FastConsensus and ECG, with FastEnsemble having better accuracy for networks with moderate to high mixing parameters, and FastEnsemble or ECG sometimes (but not always) being more accurate for the networks with low mixing parameters.

\subsection*{When the network has only partial community structure}

In Experiment 5, we examined networks that were at most half covered by clusters, and the other half was an Erd\H{o}s-R\'enyi network.
To enable a comparison to ECG and FastConsensus, we used FastEnsemble with Leiden-mod (i.e., Leiden optimizing for modularity).

In general,  the consensus clustering methods had better accuracy than Leiden-mod.
The comparison between FastEnsemble, ECG, and FastConsensus showed that no method was reliably more accurate than any other.
Examining the networks used in this experiment, we note that the Erd\H{o}s-R\'eyni graphs have mixing parameters that are at least $0.625$  and the other networks have mixing parameters that are also relatively large (i.e., at least $0.4$);
these values are perhaps large enough for us to predict, based on Experiments 1--4, that FastEnsemble should be more accurate than ECG and FastConsensus.
The trends  on the
 Erd\H{o}s-R\'enyi networks follow the predictions from the previous experiments, in that FastEnsemble was more accurate than both ECG and FastConsensus.
 However,  the relative accuracy for the networks that combine Erd\H{o}s-R\'enyi networks with either LFR or ring-of-cliques networks varied, and FastEnsemble no longer dominated the other methods.

Therefore, these trends are not obviously consistent with the trends observed in Experiments 1--4.
Moreover, both absolute and relative accuracy change with the density of the  Erd\H{o}s-R\'enyi subnetwork.

A careful examination of the mixing parameter calculation and per-node values  (see Figs E--I in S1 Appendix) is helpful in understanding this difference.
To calculate the mixing parameter of a network, the mixing parameters for each of the nodes are averaged.
The  nodes in the Erd\H{o}s-R\'enyi network that are not isolated nodes each have mixing parameter $1.0$, while the isolated nodes have mixing parameter $0.0$.
The   nodes in the  clusterable subnetworks that are paired with the Erd\H{o}s-R\'enyi networks mostly have very low mixing parameters, and average less than $0.15$.
As a result, the average mixing parameter for the networks that are formed by pairing an Erd\H{o}s-R\'enyi network with an LFR network or a ring-of-cliques network are in the moderate to high range of $0.4$ to $0.57$, but the distribution of mixing parameters is {\em bimodal}: between 50\% and 81\% are small   and the remaining ones are all maximally large at $1.0$ (see Fig I in S1 Appendix).

This is a very different kind of distribution than we have for the networks studied in Experiments 1--4 (see Figs E--H in S1 Appendix), which are for networks that have all or nearly all the nodes within communities.
For the networks in Experiments 1--4,   the per-node mixing parameters are typically concentrated around the mean with low variance (e.g., Experiments 2 and 3, see Figs F and G in S1 Appendix), and even if they  have wide variance (e.g., Experiment 1, see Fig  E in S1 Appendix),  they are nevertheless not bimodal.

\subsection*{Strict Consensus}

The Strict Consensus  is FastEnsemble with the threshold $t$ set to $1.0$; we studied two versions that differ only in how many partitions are used.
Because $t=1$ in the Strict Consensus,  unless a pair of nodes are co-clustered in every partition, the weight on the edge will be $0$; thus, the Strict Consensus variants are designed to be very  conservative.
The Strict Consensus was  explored in two experiments: Experiment 4, which examined networks that presented a challenge for the resolution limit, and Experiment 5, which examined networks that had only partial community structure.
In these experiments, the Strict Consensus had very good accuracy, including the best accuracy of all methods on the Experiment 4 networks and on the Erd\H{o}s-R\'eyni networks from Experiment 5.  This is not surprising.
We also observed that the Strict Consensus was among the better methods  for the combination of Erd\H{o}s-R\'eyni graphs with other subnetworks in Experiment 5.

Taken together, these results suggest that the Strict Consensus is capable of producing highly accurate clusterings under a range of conditions that emphasize avoiding false discovery of clusters, whether because the network  is only partially covered by communities or because it presents the resolution limit challenge.


\subsection*{Comparing ECG and FastEnsemble}
Given the algorithmic similarities between ECG and FastEnsemble, the variations in accuracy under certain conditions—sometimes favoring ECG and other times FastEnsemble—are particularly interesting.
One possible explanation is that FastEnsemble, by default, employs Leiden-mod, whereas ECG uses Louvain.
However,  as shown in Figs J--L in S1 Appendix, there is no difference in accuracy between FastEnsemble used with Louvain and FastEnsemble used with Leiden-mod for the Algorithm Design datasets, the ring-of-cliques networks, and the tree-of-cliques networks.
This suggests that the difference in accuracy between FastEnsemble and ECG is not a result of the choice between Leiden-mod and Louvain.
A more notable distinction is ECG’s reliance on 2-core-based edge weighting, which FastEnsemble does not require. Further research is needed to better understand these differences and their impact on the relative performance of these methods.

\subsection*{Computational performance}
By design, the three consensus methods we explored
(ECG, FastEnsemble, and FastConsensus) are  slower than their base methods.
Therefore, the focus here is on the relative computational performance of the three methods, as well as scalability to large networks.

The major observation is that
FastConsensus was the most computationally   expensive method: it failed to converge in 14 out of the 45 model conditions in Experiment 1 (Table A in S1 Appendix) within the allotted four-hour time limit, including on all networks with 100K nodes.
FastConsensus also failed to complete within the allowed 48 hours on three of the five Experiment 2 networks (i.e., the LFR networks based on clustered real-world networks from \cite{park2024well-journal}), while the other methods succeeded in completing on all five.
On those networks where FastConsensus was able to converge, it was much slower than both ECG and FastEnsemble, especially on the networks with more than $1,000,000$ nodes.

A comparison between ECG and FastEnsemble(Leiden-mod)  shows that neither is consistently faster than the other. f Furthermore, both completed within the allowed time on every network we explored.
On the five large LFR networks studied in Experiment 2 (taken from \cite{park2024well-journal}) that range from $\sim$ 35K to $\sim$ 3.77M nodes,  FastEnsemble is slower on three networks and faster on two (Table B, S1 Appendix).
However,  a closer analysis shows that ECG required up to 36 hours on these networks, while FastEnsemble finished within 28 hours on each of the networks.
In addition, the two methods are reasonably close in runtime on four of the five networks, and only far apart on one: the LFR cit\_patents network, which has $\sim$ 3.77 million nodes and is the largest of the networks we explored.
On that network, ECG uses 36h and 7m, while FastEnsemble uses 23h 2m.
Thus, although neither dominates the other, these preliminary results suggest that possibly FastEnsemble may have an advantage for runtime.

Finally, the runtime of FastEnsemble used with Leiden-CPM is worth examining, even though a comparison cannot be made to either ECG or FastConsensus (which can only be used with modularity optimization).
On the 22 LFR networks from Experiment 3 (which are based on Leiden-CPM clusterings of 5 real-world networks, and range up to $\sim$ 3.77 million nodes), FastEnsemble finishes in under 3 hours on every network (Table C, S1 Appendix).
This reduced runtime, compared to when FastEnsemble was used with Leiden-mod, is likely due to Leiden-CPM being faster on these networks than Leiden-mod.

\section*{Conclusions}
\label{sec:conclusions}

This study introduced FastEnsemble, a new consensus clustering method that can be used with Louvain or Leiden for optimizing  modularity or with Leidenfor optimizing under the constant Potts model.
Our study using a wide range of synthetic networks showed that FastEnsemble generally  matches  or improves the accuracy of its base method.
We also established that FastEnsemble is fast enough to use  on large networks, with more than 3 million nodes.

The comparison between FastEnsemble and two established consensus methods, ECG and FastConsensus, shows that only FastEnsemble and ECG are able to run on large networks within reasonable timeframes (e.g., 48 hours).
The relative accuracy of the three methods depends on the network and its community structure, so that no method outperforms the others under all conditions.
However, results on networks that are entirely covered by communities suggest that
ECG and/or FastConsensus may be more accurate than FastEnsemble when the mixing parameter is very low, perhaps at most $0.3$, while FastEnsemble may be reliably more accurate when the mixing parameter is larger than this value.
Results on networks that are at most half covered by communities show somewhat different trends, so that the mixing parameter is no longer a predictor of relative accuracy, and requires further investigation.

This study leaves much for future work.
First, the main focus of this study was using consensus methods for modularity optimization, but our study also explored (in Experiment 3)  using FastEnsemble  with Leiden-CPM.
Given its speed and good accuracy in that experiment,  additional investigation into the potential for this approach is merited.
 FastEnsemble also needs to be compared to   new consensus clustering methods \cite{hussain2025parallel,morea2024enhancing}, both with respect to accuracy and computational performance.

The difference in trends for clustering networks that are entirely covered by communities and those that are only partially covered by communities indicates the need to explore accuracy on a wider range of synthetic networks.
This difference could be particularly relevant to real-world networks, as some studies have argued that real-world networks are not entirely covered by communities \cite{park2024well-journal,abcdo,miasnikof2023statistical}.
Given the difficulty in knowing the ground truth community structure in real-world networks, synthetic network generators that are designed to produce networks with only partial community structure are needed.
ABCD+o \cite{abcdo}, RECCS \cite{anne2025reccs}, and EC-SBM \cite{vu2025ec} are network simulators that explicitly allow for outliers (i.e., nodes that are not in any non-singleton community) and aim to produce realistic simulated networks.
Thus, future work should examine clustering accuracy using these simulators.

To ensure a fair comparison with other consensus methods, our study focused solely on a version of FastEnsemble that employs multiple runs of a \textit{single} algorithm. We did not investigate the advanced version that enables the combination of different clustering algorithms and multi-resolution ensemble clustering. Future research should explore this functionality to determine the conditions under which combining different algorithms (e.g., Leiden-mod and Leiden-CPM) yields superior performance compared to multiple runs of each algorithm individually. Additionally, further studies should examine a broader range of networks and algorithmic combinations to gain deeper insights into these trends and identify potential variants of FastEnsemble that may achieve higher accuracy than the current default version.

Finally, to better evaluate scalability, real-world networks should be explored, especially large real-world datasets, such as the Open Citations network with approximately 75 million nodes \cite{park2024well-journal}.

\section*{Supporting information}
\textbf{S1 Appendix. Supplementary materials document.} This PDF document contains additional details about the data generation, commands for running software, and additional results provided in 3 supplementary tables and 12 supplementary figures.

\section*{Author Contributions}
\setstretch{1.5}

\textbf{Conceptualization}: Yasamin Tabatabaee, Eleanor Wedell, Tandy Warnow.

\noindent
\textbf{Data curation}: Yasamin Tabatabaee.

\noindent
\textbf{Formal analysis}: Yasamin Tabatabaee, Eleanor Wedell, Minhyuk Park

\noindent
\textbf{Funding acquisition}: Tandy Warnow

\noindent
\textbf{Investigation}: Yasamin Tabatabaee, Eleanor Wedell, Minhyuk Park, Tandy Warnow

\noindent
\textbf{Methodology}: Yasamin Tabatabaee, Eleanor Wedell, Tandy Warnow

\noindent
\textbf{Project administration}: Tandy Warnow

\noindent
\textbf{Resources}: Tandy Warnow

\noindent
\textbf{Software}: Yasamin Tabatabaee, Minhyuk Park, Eleanor Wedell

\noindent
\textbf{Supervision}: Tandy Warnow

\noindent
\textbf{Validation}: Yasamin Tabatabaee

\noindent
\textbf{Writing – original draft}:  Yasamin Tabatabaee

\noindent
\textbf{Writing – review \& editing}:   Tandy Warnow

\setstretch{1}

\section*{Acknowledgments}
This work was supported in part by a Dissertation Completion Fellowship from the Graduate College of the University of Illinois Urbana-Champaign to YT and
 by the Grainger Foundation Breakthroughs Initiative gift to the University of Illinois Urbana-Champaign to TW.

\section*{Data and Code Availability}
The code and scripts used in this study are available at https://github.com/ytabatabaee/fast-ensemble. The data are available
at https://github.com/ytabatabaee/ensemble-clustering-data.

\nolinenumbers

%
%
%
\bibliography{ref}

\begin{thebibliography}{10}

\bibitem{park2024well-journal}
Park M, Tabatabaee Y, Ramavarapu V, Liu B, Pailodi VK, Ramachandran R, et~al.
\newblock Well-connectedness and community detection.
\newblock PLOS Complex Systems. 2024;1(3):e0000009.

\bibitem{yang2016comparative}
Yang Z, Algesheimer R, Tessone CJ.
\newblock {A comparative analysis of community detection algorithms on artificial networks}.
\newblock Scientific Reports. 2016;6(1):1--18.

\bibitem{fortunato2010community}
Fortunato S.
\newblock Community detection in graphs.
\newblock Physics Reports. 2010;486(3-5):75--174.

\bibitem{newman2004finding}
Newman ME, Girvan M.
\newblock Finding and evaluating community structure in networks.
\newblock Physical Review E. 2004;69(2):026113.

\bibitem{ronhovde2010local}
Ronhovde P, Nussinov Z.
\newblock Local resolution-limit-free Potts model for community detection.
\newblock Physical Review E. 2010;81(4):046114.

\bibitem{lancichinetti2012consensus}
Lancichinetti A, Fortunato S.
\newblock Consensus clustering in complex networks.
\newblock Scientific Reports. 2012;2(1):1--7.

\bibitem{morea2024enhancing}
Morea F, De~Stefano D.
\newblock Enhancing stability and assessing uncertainty in community detection through a consensus-based approach.
\newblock arXiv. 2024;doi:{10.48550/arXiv.2408.02959}.

\bibitem{traag2019louvain}
Traag VA, Waltman L, Van~Eck NJ.
\newblock {From Louvain to Leiden: guaranteeing well-connected communities}.
\newblock Scientific Reports. 2019;9(1):1--12.

\bibitem{boyack2022improved}
Boyack KW, Klavans R.
\newblock An improved practical approach to forecasting exceptional growth in research.
\newblock Quantitative Science Studies. 2022; p. 1--25.

\bibitem{wedell2022center}
Wedell E, Park M, Korobskiy D, Warnow T, Chacko G.
\newblock Center--periphery structure in research communities.
\newblock Quantitative Science Studies. 2022;3(1):289--314.

\bibitem{strehl2002cluster}
Strehl A, Ghosh J.
\newblock Cluster ensembles---a knowledge reuse framework for combining multiple partitions.
\newblock Journal of Machine Learning Research. 2002;3(Dec):583--617.

\bibitem{tandon2019fast}
Tandon A, Albeshri A, Thayananthan V, Alhalabi W, Fortunato S.
\newblock Fast consensus clustering in complex networks.
\newblock Physical Review E. 2019;99(4):042301.

\bibitem{jeub2018multiresolution}
Jeub LG, Sporns O, Fortunato S.
\newblock Multiresolution consensus clustering in networks.
\newblock Scientific Reports. 2018;8(1):1--16.

\bibitem{goder2008consensus}
Goder A, Filkov V.
\newblock Consensus clustering algorithms: Comparison and refinement.
\newblock In: 2008 Proceedings of the Tenth Workshop on Algorithm Engineering and Experiments (ALENEX). SIAM; 2008. p. 109--117.

\bibitem{li2008weighted}
Li T, Ding C.
\newblock Weighted consensus clustering.
\newblock In: Proceedings of the 2008 SIAM International Conference on Data Mining. SIAM; 2008. p. 798--809.

\bibitem{lock2013bayesian}
Lock EF, Dunson DB.
\newblock Bayesian consensus clustering.
\newblock Bioinformatics. 2013;29(20):2610--2616.

\bibitem{van2022fast}
van Dongen S.
\newblock Fast multi-resolution consensus clustering.
\newblock bioRxiv. 2022; p. 2022--10.
\newblock doi:{10.1101/2022.10.09.511493}.

\bibitem{zhang2014scalable}
Zhang P, Moore C.
\newblock Scalable detection of statistically significant communities and hierarchies, using message passing for modularity.
\newblock Proceedings of the National Academy of Sciences. 2014;111(51):18144--18149.

\bibitem{hussain2025parallel}
Hussain MT, Halappanavar M, Chatterjee S, Radicchi F, Fortunato S, Azad A.
\newblock Parallel median consensus clustering in complex networks.
\newblock Scientific Reports. 2025;15(1):3788.

\bibitem{de2011generalized}
De~Meo P, Ferrara E, Fiumara G, Provetti A.
\newblock Generalized {louvain} method for community detection in large networks.
\newblock In: 2011 11th international conference on intelligent systems design and applications. IEEE; 2011. p. 88--93.

\bibitem{que2015scalable-louvain}
Que X, Checconi F, Petrini F, Gunnels JA.
\newblock Scalable community detection with the {Louvain} algorithm.
\newblock In: 2015 IEEE International Parallel and Distributed Processing Symposium. IEEE; 2015. p. 28--37.

\bibitem{poulin2019ensemble}
Poulin V, Th{\'e}berge F.
\newblock Ensemble clustering for graphs: comparisons and applications.
\newblock Applied Network Science. 2019;4(1):51.

\bibitem{tabatabaee2024fastensemble}
Tabatabaee Y, Wedell E, Park M, Warnow T.
\newblock {FastEnsemble:} A new scalable ensemble clustering method.
\newblock arXiv. 2024;doi:{10.48550/arXiv.2409.02077}.

\bibitem{lancichinetti2008benchmark}
Lancichinetti A, Fortunato S, Radicchi F.
\newblock Benchmark graphs for testing community detection algorithms.
\newblock Physical Review E. 2008;78(4):046110.

\bibitem{lfr-generation-code}
Tabatabaee Y. Real network emulation using {LFR} graphs; 2023.
\newblock \url{https://github.com/ytabatabaee/emulate-real-nets}.

\bibitem{fortunato2007resolution}
Fortunato S, Barthelemy M.
\newblock Resolution limit in community detection.
\newblock Proceedings of the {N}ational {A}cademy of {S}ciences. 2007;104(1):36--41.

\bibitem{newman2003mixing}
Newman ME.
\newblock Mixing patterns in networks.
\newblock Physical Review E. 2003;67(2):026126.

\bibitem{jiang2020community}
Jiang H, Liu Z, Liu C, Su Y, Zhang X.
\newblock Community detection in complex networks with an ambiguous structure using central node based link prediction.
\newblock Knowledge-Based Systems. 2020;195:105626.

\bibitem{erdos-renyi}
Erdős P, Rényi A.
\newblock On Random Graphs.
\newblock Publicationes Mathematicae. 1959;6(3-4):290–297.

\bibitem{pedregosa2011scikit}
Pedregosa F, Varoquaux G, Gramfort A, Michel V, Thirion B, Grisel O, et~al.
\newblock Scikit-learn: Machine learning in Python.
\newblock the Journal of Machine Learning Research. 2011;12:2825--2830.

\bibitem{traag2011narrow}
Traag VA, Van~Dooren P, Nesterov Y.
\newblock Narrow scope for resolution-limit-free community detection.
\newblock Physical Review E. 2011;84(1):016114.

\bibitem{lancichinetti2011limits}
Lancichinetti A, Fortunato S.
\newblock Limits of modularity maximization in community detection.
\newblock Physical Review E—Statistical, Nonlinear, and Soft Matter Physics. 2011;84(6):066122.

\bibitem{poulin2019ECG}
Poulin V, Th{\'e}berge F.
\newblock Ensemble clustering for graphs.
\newblock In: Complex Networks and Their Applications VII: Volume 1 Proceedings The 7th International Conference on Complex Networks and Their Applications COMPLEX NETWORKS 2018 7. Springer; 2019. p. 231--243.

\bibitem{abcdo}
Kami{\'n}ski B, Pra{\l}at P, Th{\'e}berge F.
\newblock Artificial benchmark for community detection with outliers {(ABCD+o)}.
\newblock Applied Network Science. 2023;8(1):25.

\bibitem{miasnikof2023statistical}
Miasnikof P, Shestopaloff AY, Raigorodskii A.
\newblock Statistical power, accuracy, reproducibility and robustness of a graph clusterability test.
\newblock International Journal of Data Science and Analytics. 2023;15(4):379--390.

\bibitem{anne2025reccs}
Anne L, Vu-Le TA, Park M, Warnow T, Chacko G.
\newblock RECCS: Realistic Cluster Connectivity Simulator for Synthetic Network Generation.
\newblock arXiv preprint arXiv:250202050. 2025;doi:{10.48550/arXiv.2502.02050}.

\bibitem{vu2025ec}
Vu-Le TA, Anne L, Chacko G, Warnow T.
\newblock EC-SBM Synthetic Network Generator.
\newblock arXiv. 2025;doi:{10.48550/arXiv.2502.03662}.


\end{thebibliography}


\begin{thebibliography}{}

\bibitem[Fortunato, 2025]{fortunato-resources}
Fortunato, S. (2025).
\newblock Resources.
\newblock {https://www.santofortunato.net/resources}.

\bibitem[Hagberg et~al., 2008]{hagberg2008exploring-conf}
Hagberg, A., Swart, P., and S~Chult, D. (2008).
\newblock Exploring network structure, dynamics, and function using {NetworkX}.
\newblock {\em 7th Python in Science Conference (SciPy2008)}, pages 11--15.
\newblock \url{http://conference.scipy.org/proceedings/scipy2008/}.

\bibitem[Lancichinetti et~al., 2008]{lancichinetti2008benchmark}
Lancichinetti, A., Fortunato, S., and Radicchi, F. (2008).
\newblock Benchmark graphs for testing community detection algorithms.
\newblock {\em Physical Review E}, 78(4):046110.

\bibitem[Park et~al., 2024]{park2024well-journal}
Park, M., Tabatabaee, Y., Ramavarapu, V., Liu, B., Pailodi, V.~K., Ramachandran, R., Korobskiy, D., Ayres, F., Chacko, G., and Warnow, T. (2024).
\newblock Well-connectedness and community detection.
\newblock {\em PLOS Complex Systems}, 1(3):e0000009.

\bibitem[Poulin and Th{\'e}berge, 2019]{poulin2019ensemble}
Poulin, V. and Th{\'e}berge, F. (2019).
\newblock Ensemble clustering for graphs: comparisons and applications.
\newblock {\em Applied Network Science}, 4(1):51.

\bibitem[Tabatabaee, 2023]{tabatabaee2023improving}
Tabatabaee, Y. (2023).
\newblock Improving the accuracy of community detection methods using {Connectivity Modifier}.
\newblock MS Thesis, Department of Computer Science, University of Illinois Urbana-Champaign.

\bibitem[Tandon et~al., 2019]{tandon2019fast}
Tandon, A., Albeshri, A., Thayananthan, V., Alhalabi, W., and Fortunato, S. (2019).
\newblock Fast consensus clustering in complex networks.
\newblock {\em Physical Review E}, 99(4):042301.

\end{thebibliography}

\end{document}


\maketitle

\tableofcontents
\listoftables
\listoffigures
\newpage


\section{Details of the Experimental Study}
\subsection{Simulated Datasets}
\subsubsection{Algorithm design datasets}

We used the library NetworkX \citep{hagberg2008exploring-conf} with the following command to generate the LFR graphs used in the algorithm design experiment:
\begin{lstlisting}[basicstyle=\ttfamily\small]
graph = nx.generators.community.LFR_benchmark_graph(n=n, tau1=3, tau2=1.5,
mu=mu, average_degree=d, min_community=10, seed=1932)
\end{lstlisting} 
where the mixing parameter \texttt{mu} varies between 0.1 and 0.9, the average degree \texttt{d} varies between 5, 10 and 20, and the number \texttt{n} of nodes in the network varies between 1000 to 100,000.
The parameters $\tau_1$ and $\tau_2$ are exponents for the degree and community size distributions respectively and $c_{min}$ is the minimum community size.
This collection of networks were used in Experiment 1.

In addition, we attempted to regenerate the LFR datasets used in Figure 2 of \cite{tandon2019fast} study. 
We were not able to re-generate these graphs using NetworkX, potentially due to the choice of parameters. Therefore we used the original implementation of LFR software in c++ available at \cite{fortunato-resources} to generate these networks with the following command:
\begin{lstlisting}[basicstyle=\ttfamily\small]
./binary_networks/benchmark -N 10000 -k 20 -maxk 50 -mu <mixing-parameter>
-maxc 100 -minc 10 -t1 2 -t2 3
\end{lstlisting} 

Note that \cite{lancichinetti2008benchmark} suggest that the normal range for parameter $\tau_2$ is 1 to 2, while \cite{tandon2019fast} set this value  to 3.
It seems possible that this discrepancy may have
resulted in problems in generating these set of networks using NetworkX. 
Since the parameters $\tau_1$ and $\tau_2$ used in our simulations for Figure 1 are in the range suggested by \cite{lancichinetti2008benchmark} while the parameters used in  \cite{tandon2019fast} are not,  we consider the LFR networks we used in Experiment 1 to be preferable.


\subsubsection{Synthetic datasets derived from real-world networks}

For Experiments 2 and 3, we used the 34 LFR networks from \cite{park2024well-journal} that were generated based on the properties of six real-world networks and their Leiden clusterings, optimizing respectively for modularity or CPM with different resolution values (refer to \cite{park2024well-journal} for further information about network generation protocol).  We omit those LFR graphs that had high proportion of disconnected ground-truth communities, therefore using only 27 out of 34 graphs.
These are freely available in the Illinois Data Bank at \url{https://doi.org/10.13012/B2IDB-6271968_V1}.

\subsubsection{Ring-of-cliques and Tree-of-cliques networks}
\label{sec:ring}

 Experiment 4 explores accuracy on ring-of-cliques networks that are formed by arranging $n$ cliques in a ring, with each ring of size $10$. We used the following command to generate these networks with the software NetworkX, where \texttt{n} and \texttt{k} determine the number of cliques and size of each clique, respectively.
\begin{lstlisting}[basicstyle=\ttfamily\small]
nx.ring_of_cliques(num_cliques=n, clique_size=k)
\end{lstlisting} 

\noindent
To create Tree-of-Cliques networks, we used the following custom code
\begin{lstlisting}[basicstyle=\ttfamily\small]
def gen_tree_of_cliques(k, n):
  cliques = [nx.complete_graph(k) for _ in range(n)]
  tree = nx.random_tree(n)
  tree_of_cliques = nx.disjoint_union_all(cliques)
  for s, d in tree.edges():
    tree_of_cliques.add_edge(s*k+k-1, d*k)
  return tree_of_cliques
\end{lstlisting} 
that first generates a random tree of size \texttt{n} and \texttt{n} disjoint cliques of size \texttt{k}, and then connects the cliques according to the structure of the tree, so that for each edge \texttt{(s,d)} in the tree, the last node of clique \texttt{s} is connected to the first node of clique \texttt{d}.

\subsubsection{Synthetic datasets composed of Erd\H{o}s-R\'enyi graphs}

The datasets we use in Experiment 5 are either Erd\H{o}s-R\'enyi graphs or are composed of a combination of Erd\H{o}s-R\'enyi (ER) graphs with an LFR graph or a ring-of-cliques.

The Erd\H{o}s-R\'enyi graphs have 1000 nodes.
We used NetworkX to generate these Erd\H{o}s-R\'enyi graphs, with the following command, where \texttt{p} specifies the probability of creating an edge that affects the density of the graph:
\begin{lstlisting}[basicstyle=\ttfamily\small]
graph = nx.erdos_renyi_graph(n=1000, p=p)
\end{lstlisting} 

To combine these graphs with synthetic graphs with known community structure, we created an LFR graph with 1000 nodes using NetworkX with the following command 
\begin{lstlisting}[basicstyle=\ttfamily\small]
lfr = nx.generators.community.LFR_benchmark_graph(n=1000, tau1=3, tau2=1.5, mu=0.1, average_degree=9.198, min_community=45, seed=10)
\end{lstlisting} 
This graph has 14 ground-truth communities with sizes [45, 47, 57, 59, 60, 61, 70, 74, 74, 87, 88, 91, 91, 96].

To attach this graph to the  Erd\H{o}s-R\'enyi graphs with various densities, we used the following commands:
\begin{lstlisting}[basicstyle=\ttfamily\small]
graph = nx.erdos_renyi_graph(n=1000, p=p)
graph = nx.disjoint_union_all([graph, lfr])
graph.add_edge(0, graph.number_of_nodes() - 1)
\end{lstlisting} 
\noindent
that creates an Erd\H{o}s-R\'enyi graph using NetworkX with edge probability $p$ and connects it to the LFR graph created before using a single edge.
The ground-truth community structure for the ER-LFR graphs is assumed to be the 14 communities in the LFR graph in addition to 1000 singletons for the ER graph. 

Similarly, we created combinations of Erd\H{o}s-R\'enyi graphs of various densities with a Ring-of-Cliques networks with 100 cliques of size 10 each (refer to Section \ref{sec:ring} for the command for generating Ring-of-Cliques graphs). 

Note that these Erd\H{o}s-R\'enyi networks can have isolated nodes, and hence the edge list representation for these networks may suggest fewer than 1000 nodes; the full synthetic network for the Erd\H{o}s-R\'enyi graph, however, has 1000 nodes.

\clearpage
\subsection{Methods and Software Commands}

\paragraph{FastEnsemble} The code for FastEnsemble is available at \url{https://github.com/ytabatabaee/fast-ensemble/blob/main/fast_ensemble.py}. 
The following command can be used to run it, where \texttt{-t} refers to the threshold for removing weak edges and \texttt{-p} specifies the number of partitions. The flag \texttt{--noweight} specifies that the method should ignore edge weights when clustering.
In our experiments, we ran FastEnsemble with Leiden-mod and Louvain for modularity clustering without the \texttt{--noweight} flag.
When we used FastEnsemble with Leiden-CPM we used it with the \texttt{--noweight} flag.

\begin{lstlisting}[language=python]
python <git root>/fast_ensemble.py -n <edgelist> -t <threshold> -alg <algorithm> [-r <resolution-value>] -p <number-of-partitions> [--noweight]
\end{lstlisting}
Commit id used: 0cce8ce


\paragraph{FastConsensus.} The FastConsensus \citep{tandon2019fast} software is available at \url{https://github.com/kaiser-dan/fastconsensus}. We used the following command to run it
\begin{lstlisting}
python <git root>/fast_consensus.py -f <Input network> --alg louvain -np 10 -t 0.2 -d 0.02
\end{lstlisting}

\noindent
where \texttt{-np} specifies the number of partitions, \texttt{-t} specifies the threshold for removing weak edges, \texttt{-d} is the convergence threshold and \texttt{--alg} specifies the clustering algorithm which is by default Louvain. 
Commit id used: 9bf993b


\paragraph{ECG.} The software for ECG \citep{poulin2019ensemble} is available at \url{https://github.com/ftheberge/Ensemble-Clustering-for-Graphs/tree/master}. We wrote a custom script to run it, which is available at  \url{https://github.com/ytabatabaee/fast-ensemble/blob/main/scripts/ECG.py}. We used the following command to run ECG using this script:

\begin{lstlisting}
python <git root>/ECG.py <Input network> <Output file>
\end{lstlisting}
Commit id used: ce21601 with igraph version 0.9.7 on Python 3.9.18



\paragraph{Clustering Accuracy Evaluation} The script for calculating accuracy metrics (including NMI, AMI, ARI, FNR, FPR, precision, recall, F1-score) is available at \url{https://github.com/ytabatabaee/fast-ensemble/blob/main/scripts/clustering_accuracy.py}. We use the scikit-learn library to calculate NMI, AMI and ARI, and used a custom code to calculate the other measures. The script for calculating accuracy can be run with the following command:
\begin{lstlisting}
python <git root>/clustering_accuracy.py gt <Ground-truth membership> -p <Estimated partition>
\end{lstlisting}
Commit id used: 5ea5f66

\paragraph{Mixing parameter calculation}

The mixing parameter of a network for a given clustering is the average, across all the nodes in the network, of the ratio between number of neighbors of the node outside its community and its total degree.
Nodes that are not in any cluster (community) have a mixing parameter that depends on whether they are isolated (i.e., have no neighbors) or not. The mixing parameter for the nodes that are isolated is $0.0$; those that are not isolated have mixing parameter $1.0$.
On synthetic networks, we have the ground-truth clustering, and  the mixing parameter was calculated using the network and that ground-truth clustering. 
The script for calculating the mixing parameter of a network/clustering pair is available at \url{https://github.com/ytabatabaee/emulate-real-nets/blob/main/estimate_properties.py}.
Commit id used: 867e025


\clearpage
\section{Derivation of mixing parameter for ring-of-cliques networks}
For a ring-of-cliques network with $n$ cliques of size $k$, the mixing parameter with respect to the ground-truth can be calculated as 
\begin{equation}
    \hat{\mu} = \frac{1}{nk}\sum_{i \in \{1, \dots, nk\}}\frac{d_i^{out}}{d_i^{in} + d_i^{out}} = \frac{1}{nk} \times 2n \times \frac{1}{(k-1)+1} = \frac{2}{k^2}
\end{equation}
according to Eq 1 in main text, as $d_i^{out}$ is zero for all except the two nodes in each clique that are connected to other cliques, and these two nodes are connected to $k-1$ nodes inside their community and one node outside, and therefore $\frac{d_i^{out}}{d_i^{in} + d_i^{out}} = \frac{1}{k}$. 

\clearpage
\section{Additional Tables}

\begin{table}[ht!]
\centering
\caption[Failures to complete for FastConsensus (Experiments 1 and 2)]{Failures to complete for FastConsensus on synthetic LFR datasets from the algorithm design experiment (left) and the \cite{park2024well-journal} dataset based on modularity clusterings  (right). The time limit was set to 4 hours for the algorithm design dataset and 48 hours for the \cite{park2024well-journal} datasets.}
\begin{tabular}{|lll|}
\hline
$n$       & $d_{avg}$                    & $\mu$ \\ \hline
1,000   & \multicolumn{1}{l|}{10} &   0.9 \\
10,000  & \multicolumn{1}{l|}{5}  &   - \\
10,000  & \multicolumn{1}{l|}{10} &   0.4, 0.5, 0.8 \\
10,000  & \multicolumn{1}{l|}{20} &   0.5 \\
100,000 & \multicolumn{1}{l|}{10} &   0.1 to 0.9\\ \hline
\end{tabular}
\hspace{1cm}
\begin{tabular}{l|ll|}
\cline{2-3}
                                      & n         & m          \\ \hline
\multicolumn{1}{|l|}{CEN-mod}             & 3,000,000 & 20,821,202 \\
\multicolumn{1}{|l|}{open\_citations-mod} & 3,000,000 & 55,128,496 \\
\multicolumn{1}{|l|}{cit\_patents-mod}    & 3,774,768 & 15,648,081 \\ \hline
\end{tabular}
\label{table:failure}
\end{table}

\begin{table}[ht!]
\centering
\caption[Clustering accuracy (ARI/NMI) and runtime on simulated modularity-based LFR networks (Experiment 2)]{{Clustering accuracy (ARI/NMI) and runtime of methods on simulated modularity-based LFR networks from Experiment 2.} ``n.d." stands for no data due to method failing to return a clustering within 48 hours. 
}
\begin{tabular}{lllll}
\hline
                                       &                       & ARI    & NMI    & runtime           \\
\hline
\hline
\multirow{4}{*}{LFR cit\_hepph mod}    & FastEnsemble(default) & 0.9992 & 0.9949  & 43s                 \\
                                       & FastConsensus         & 0.9812 & 0.9947 & 21m 54s           \\
                                       & ECG                   & 1.0000 & 1.0000 & 15s               \\
                                       & Leiden-mod            & 0.9991 & 0.9947  & 3s                 \\
\hline
\multirow{4}{*}{LFR wiki\_topcats mod} & FastEnsemble(default) & 0.8486 & 0.9923 & 7h 36m 24s                 \\
                                       & FastConsensus         & 0.9540 & 0.9997 & 14h 34m 39s       \\
                                       & ECG                   & 0.0000 & 0.5767 & 6h 20m 36s        \\
                                       & Leiden-mod            & 0.0990 & 0.8252 & 1m 38s                 \\
\hline
\multirow{4}{*}{LFR cit\_patents mod}  & FastEnsemble(default) & 0.8511 & 0.9916 & 23h 1m 35s                 \\
                                       & FastConsensus         & n.d.   & n.d.   & \textgreater{}2d \\
                                       & ECG                   & 0.0000 & 0.4673 & 1d 12h 8m         \\
                                       & Leiden-mod            & 0.1374 &  0.7749 & 2m 48s                 \\
\hline
\multirow{4}{*}{LFR CEN mod}           & FastEnsemble(default) & 0.8820 & 0.9882 & 12h 8m 47s                \\
                                       & FastConsensus         & n.d.   & n.d.   & \textgreater{}2d \\
                                       & ECG                   & 0.9463 & 0.9803 & 12h 38m            \\
                                       & Leiden-mod            & 0.4141 & 0.8973 & 2m 31s                 \\
\hline
\multirow{4}{*}{LFR open\_citations mod}            & FastEnsemble(default) & 0.8145 & 0.9889 & 1d 3h 52m 6s        \\
                                       & FastConsensus         & n.d.   & n.d.   &  \textgreater{}2d \\
                                       & ECG                   & 0.0000 & 0.5013 & 21h 59m         \\
                                       & Leiden-mod            & 0.1502 & 0.8378 &   3m 37s               \\
\hline    
\end{tabular}
\end{table}

\begin{table}[ht!]
\caption[Runtime of FastEnsemble and Leiden-CPM on simulated CPM-based LFR networks (Experiment 3).]{Runtime of FastEnsemble (top) and Leiden-CPM (bottom) on simulated CPM-based LFR networks used in Experiment  3. In each case, FastEnsemble is used with Leiden-CPM with the same resolution value as was used to provide parameters to the LFR software to generate the network. Here, ``n.a." refers to conditions were results are not reported, either due to the LFR software not being able to generate a graph for that condition, or due to a high proportion of the ground-truth communities being  disconnected.
}
\centering
\begin{tabular}{l|lllll|}
\cline{2-6}
   FastEnsemble(Leiden-CPM)                                 & 0.0001     & 0.001      & 0.01       & 0.1       & 0.5        \\ \hline
\multicolumn{1}{|l|}{cit\_hepph}    & 22s        & 26s        & 30s        & 39s       & 56s        \\
\multicolumn{1}{|l|}{wiki\_topcats} & 57m 2s     & 45m 21s    & 51m 54s    & 51m 19s   & n.a.       \\
\multicolumn{1}{|l|}{cit\_patents}  & 39m 51s    & 34m 53s    & 38m 46s    & 45m 11s   & 45m 35s    \\
\multicolumn{1}{|l|}{CEN}           & 1h 12m 48s & 1h 7m 19s  & 1h 6m 7s   & n.a.       & n.a.    \\
\multicolumn{1}{|l|}{open\_citations}            & 2h 34m 58s & 1h 43m 24s & 1h 43m 51s & 2h 39m 5s & 2h 10m 57s \\
\hline
\end{tabular}
\newline
\vspace{1cm}
\newline
\begin{tabular}{l|lllll|}
\cline{2-6}
   Leiden-CPM                                & 0.0001     & 0.001      & 0.01       & 0.1       & 0.5        \\ \hline
\multicolumn{1}{|l|}{cit\_hepph}    &   20s      &     21s    &      21s   &    24s   &   23s    \\
\multicolumn{1}{|l|}{wiki\_topcats} &    11m 31s   & 12m 5s   &    12 m 42s &  13m 19s  & n.a.       \\
\multicolumn{1}{|l|}{cit\_patents}  &   9m 18s  & 9m 6s    &  10m 9s  &  10m 14s  &  9m 31s  \\
\multicolumn{1}{|l|}{CEN}           &  13m 54s &  13m 37s &  13m 25s  & n.a.       & n.a.    \\
\multicolumn{1}{|l|}{open\_citations}    & 30m 50s & 29m 41s & 26m 28s & 30m 17s & 36m 33s \\
\hline
\end{tabular}
\end{table}

\clearpage
\section{Additional Figures}

\begin{figure}[h!]
\centering
\includegraphics[width=1\textwidth]{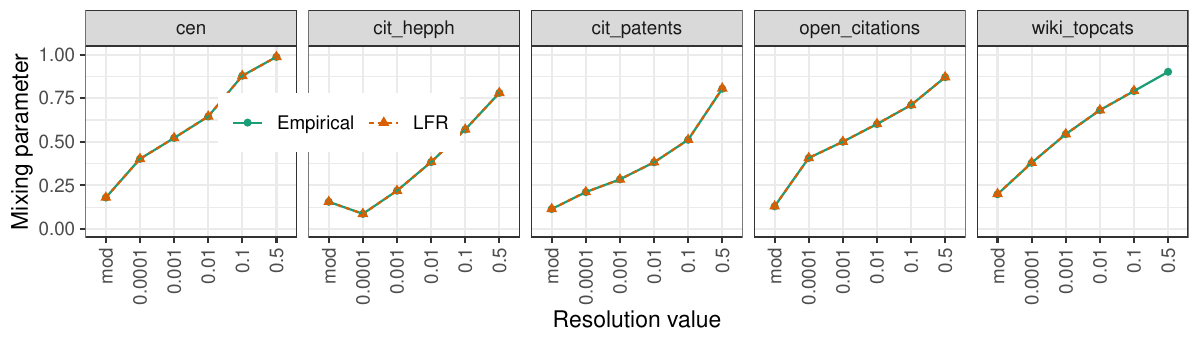}
\caption[Mixing parameters for LFR networks from \cite{park2024well-journal} (Experiments 2 and 5)]
{\textbf{Mixing parameters for LFR networks based on clustered real-world networks for Experiments 2 and 3, and their corresponding real-world networks}. 
 Each condition on the x-axis corresponds to a \emph{different} LFR network, generated based on Leiden-modularity or Leiden-CPM with that specific resolution parameter. The mixing parameter results for the real-world (empirical) networks are calculated using the specified Leiden clustering, and for the synthetic network using the LFR ground-truth community structure (the green and orange lines respectively). No results are shown for wiki\_topcats network with resolution value $r= 0.5$  due to LFR failing to generate networks for that setting.  Note that the LFR network based on Leiden-modularity clustering has the smallest mixing parameter, and that the mixing parameters for the LFR networks based on Leiden-CPM clusterings increase with the resolution value. Note also that the mixing parameters for the real-world and corresponding LFR networks are nearly identical. (Figure modified from Figure 5.2 in \cite{tabatabaee2023improving}.)
 }
\label{fig:mixing-params}
\end{figure}

\begin{figure}[h!]
    \centering  
    \includegraphics[width=0.49\textwidth]{figs/training_exp.pdf}
    \includegraphics[width=0.49\textwidth]{figs/training_threshold.pdf}
    \includegraphics[width=0.65\textwidth]{figs/training_legend.pdf}
    \includegraphics[width=1\textwidth]{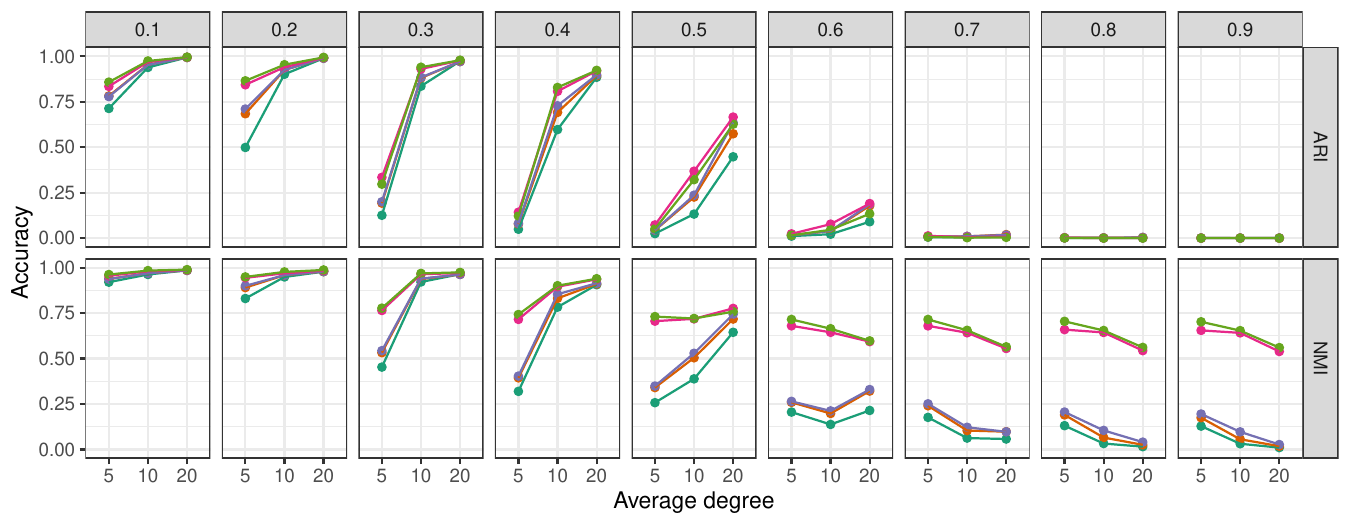}
    \includegraphics[width=1\textwidth]{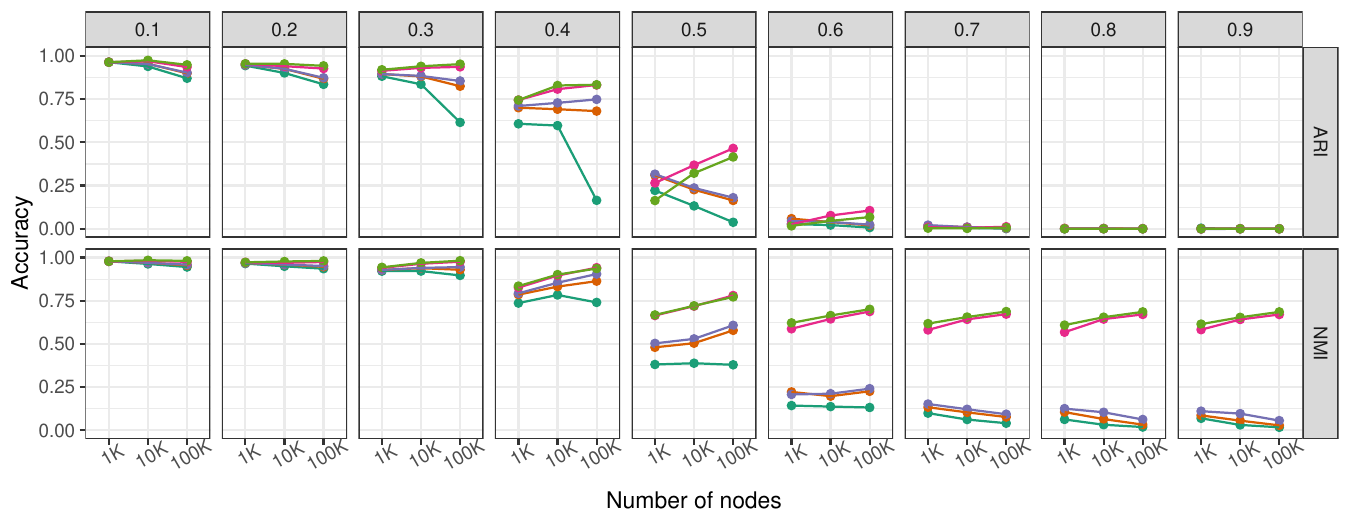}
    \caption[Experiment 1a: Setting the default for parameter $t$ in FastEnsemble]{\textbf{Experiment 1a: Setting the default value for $t$ in  FastEnsemble.} 
    Each plot shows ARI and NMI accuracy for Leiden-mod and FastEnsemble using four different threshold values on the algorithm design networks.
    Top left: Accuracy on networks with 10,000 nodes as a function of the model mixing parameter (x-axis).  Top right: Accuracy as  a function of the threshold value on the networks   with model mixing parameter $0.5$. 
    Middle: Accuracy as a function of the average node degree and model mixing parameter. 
    Bottom: Accuracy as a function of the network size.  
    The default model condition for the LFR graphs have 10,000 nodes and  average degree 10.
    FE stands for FastEnsemble.}
    \label{fig:expt1a}
\end{figure}

\begin{figure}[tbp!]
    \centering
    \includegraphics[width=1\textwidth]{figs/comparison_methods_0.8.pdf}
    \includegraphics[width=1\textwidth]{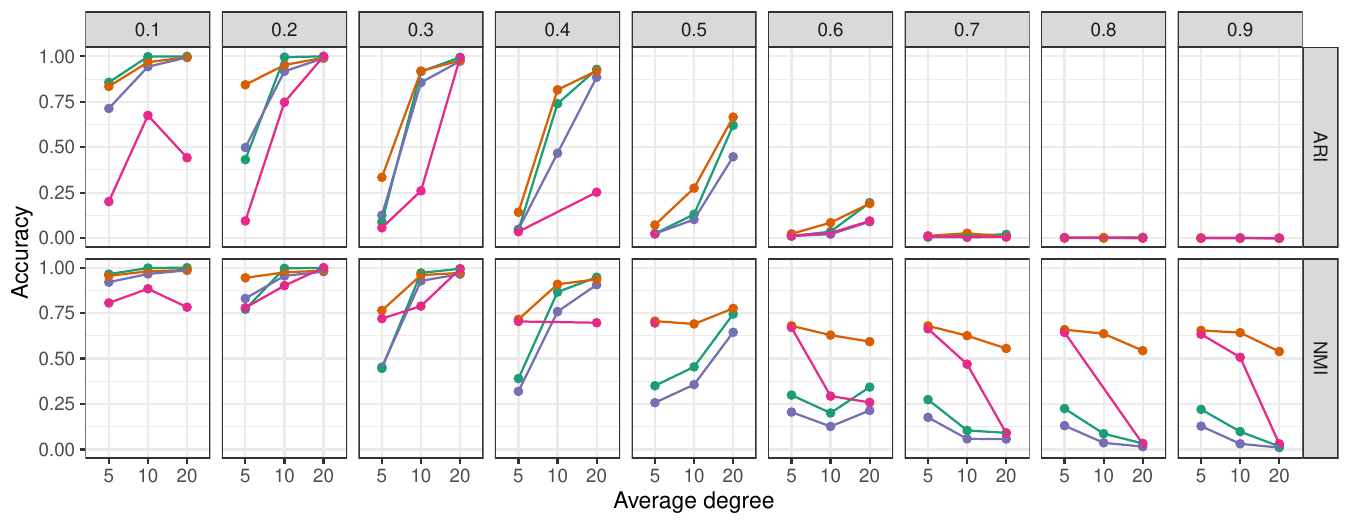}
    \includegraphics[width=1\textwidth]{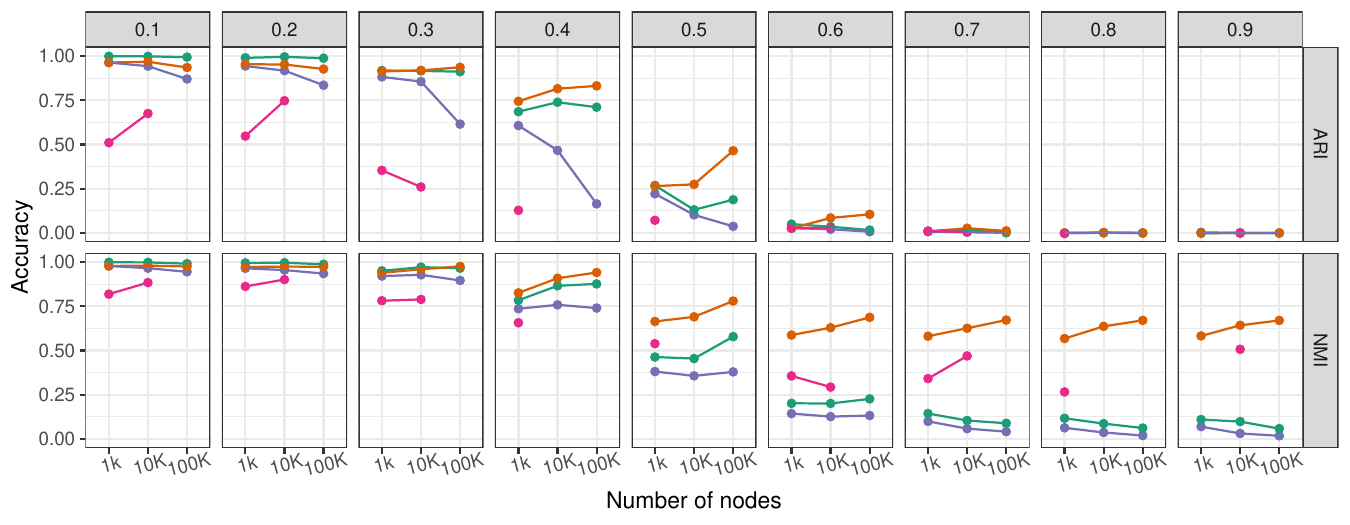}
    \caption[Experiment 1b: Evaluating consensus clustering pipelines on the algorithm design datasets]{\textbf{Experiment 1b: Evaluating modularity-based consensus clustering pipelines on algorithm design datasets.} Top: Accuracy on the algorithm design datasets  as the mixing parameter for the  network changes (values on the x-axis). Middle: Accuracy on algorithm design datasets as the average  node degree and mixing parameter changes.  Bottom: Accuracy on the algorithm design datasets as the network size and mixing parameter changes.  Note:  FastConsensus failed to converge within four hours in several model conditions (Table A in S1 Appendix), including all networks of size 100K. 
    }
    \label{fig:expt1b}
\end{figure}


\begin{figure}[h!]
    \centering
    \includegraphics[width=0.8\textwidth]{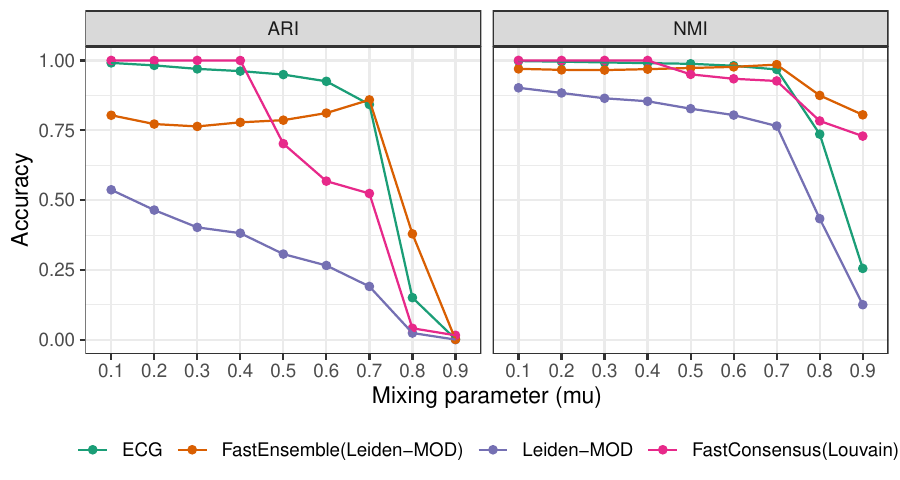}
    \caption[Results on the original \cite{tandon2019fast} networks]{\textbf{Accuracy of modularity-based clusterings on \cite{tandon2019fast} dataset.} The dataset is created based on the model conditions of Fig.~2 in \cite{tandon2019fast} with parameters $n=10,000$, degree exponents $\tau_1 = 2$, $\tau_2 = 3$, and $k_{avg}=20$, $k_{max}=50$, $c_{min} = 10$ and $c_{max} = 100$ with mixing parameter $\mu$ that varies between 0.1 to 0.9.  The methods compared are ECG, Leiden-mod, FastConsensus, and FastEnsemble(Leiden-mod). The three consensus methods have higher accuracy than Leiden-mod, but their relative accuracy  depends on the mixing parameter: FastConsensus has the best accuracy for the smallest mixing parameters, followed by ECG; FastEnsemble has the best accuracy for the largest mixing parameters.
    }
    \label{fig:expt2-training}
\end{figure}

\begin{figure}[ht!]
    \centering
    \includegraphics[width=0.6\textwidth]{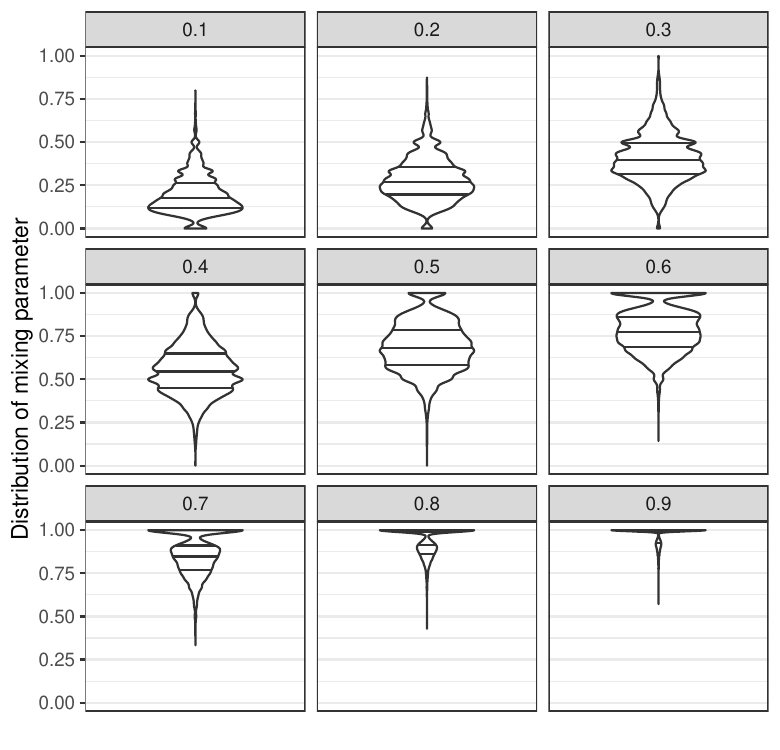}
    \caption[Distribution of the per-node mixing parameters for the networks used in Experiment 1]{\textbf{Distribution of the per-node mixing parameters for the networks used in Experiment 1.} Each panel corresponds to a \textit{model} mixing parameter used to generate the LFR network and the y-axis shows the \textit{estimated} mixing parameter. Mixing parameters are calculated with respect to the LFR ground-truth community structure. The 25\%, 50\% and 75\% quantiles are specified with black lines on the violin plots. The conditions have the 
default average degree (10) and network size (10,000).
    }
\label{fig:mu-dist-alg-design}
\end{figure}

\begin{figure}[ht!]
    \centering
    \includegraphics[width=0.7\textwidth]{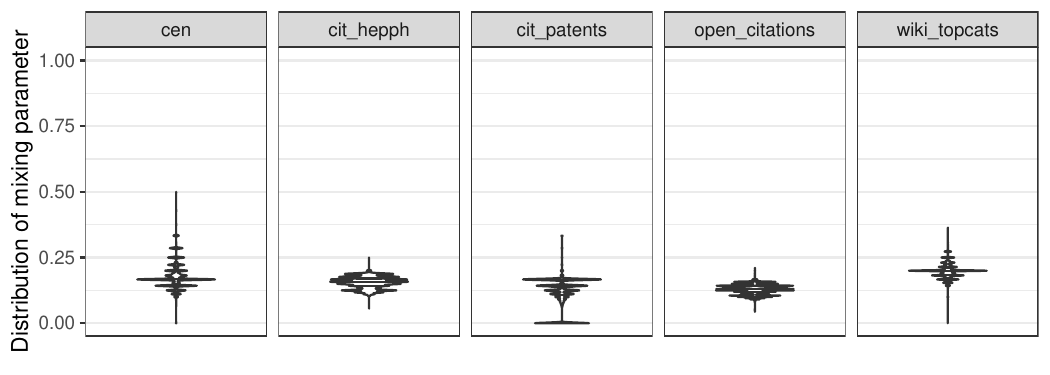}
    \caption[Distribution of the per-node mixing parameters for the networks used in Experiment 2.]{\textbf{Distribution of the per-node mixing parameters for the modularity-based LFR networks from \cite{park2024well-journal} used in Experiment 2.} The LFR networks are generated based on a Leiden-mod clustering of a corresponding empirical network (shown on the panels). Mixing parameters are calculated with respect to the LFR ground-truth community structure. 
    }
\label{fig:mu-dist-cm-mod}
\end{figure}

\begin{figure}[ht!]
    \centering
    \includegraphics[width=0.9\textwidth]{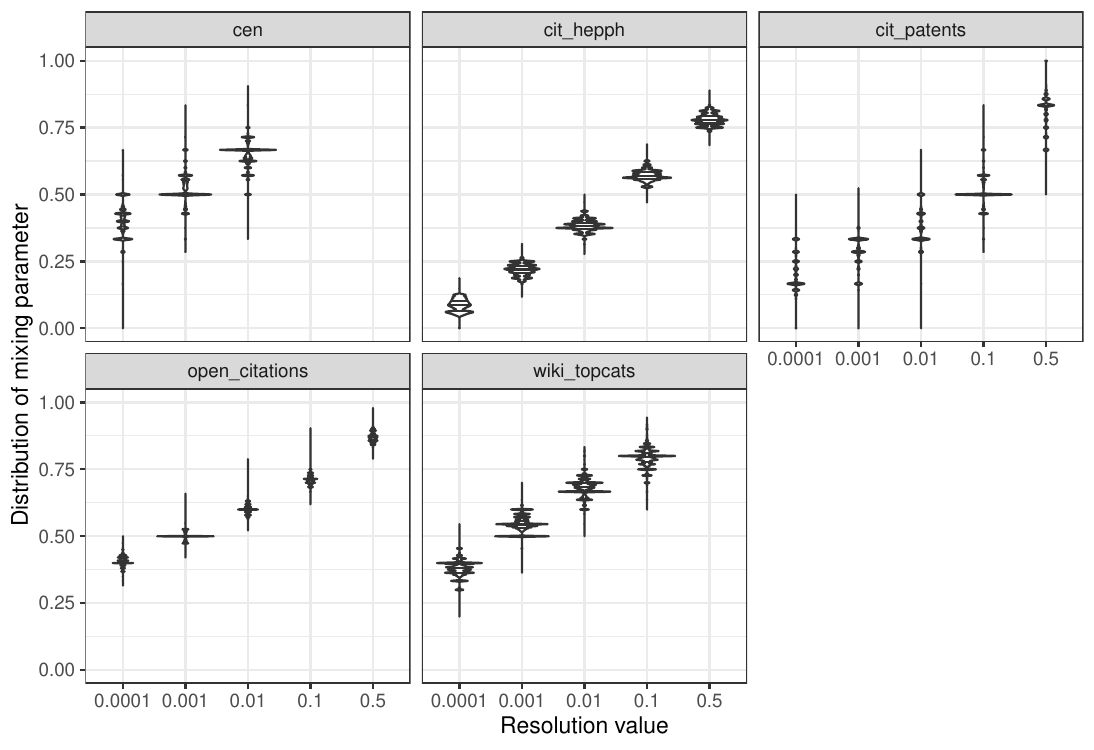}
    \caption[Distribution of the per-node mixing parameters for the  networks used in Experiment 3.]{\textbf{Distribution of the per-node mixing parameters for    networks used in Experiment 3.} Each condition on the x-axis corresponds to a \emph{different} LFR network corresponding to an empirical network (shown on the panels), generated based on Leiden-CPM with that specific resolution parameter; these networks are taken from \cite{park2024well-journal}.   Mixing parameters are calculated with respect to the LFR ground-truth community structure. 
     Results are not shown for three conditions:  LFR graphs with a large fraction of disconnected ground truth clusters (the two CEN networks) or when the LFR software failed to create a network for the provided parameters (the wiki\_topcats network). 
    }
\label{fig:mu-dist-cm-cpm}
\end{figure}

\begin{figure}[ht!]
    \centering
    \includegraphics[width=0.5\textwidth]{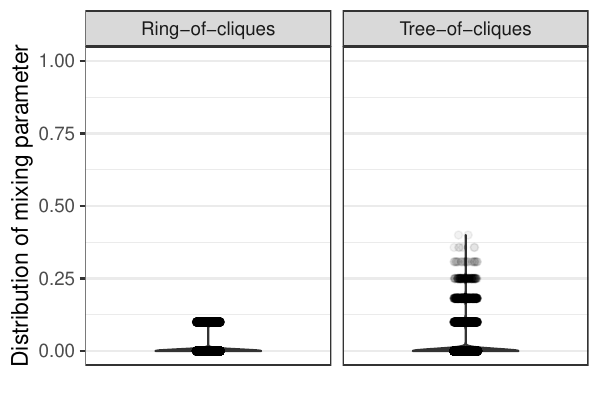}
    \caption[Distribution of the per-node mixing parameters for the   networks used in Experiment 4]{\textbf{Distribution of the per-node mixing parameters for the Ring-of-cliques and Tree-of-cliques networks used in Experiment 4.} Only one example of each network is shown, since the distributions are independent of the network size for Ring-of-cliques and Tree-of-cliques networks. The ring and tree of cliques have 1000 cliques of size 10. 
    }
\label{fig:mu-dist-trivial}
\end{figure}

\begin{figure}[ht!]
    \centering
    A) Erd\H{o}s-R\'eyni\hfill~\\\vspace{-0pt}
    \includegraphics[width=0.9\textwidth]{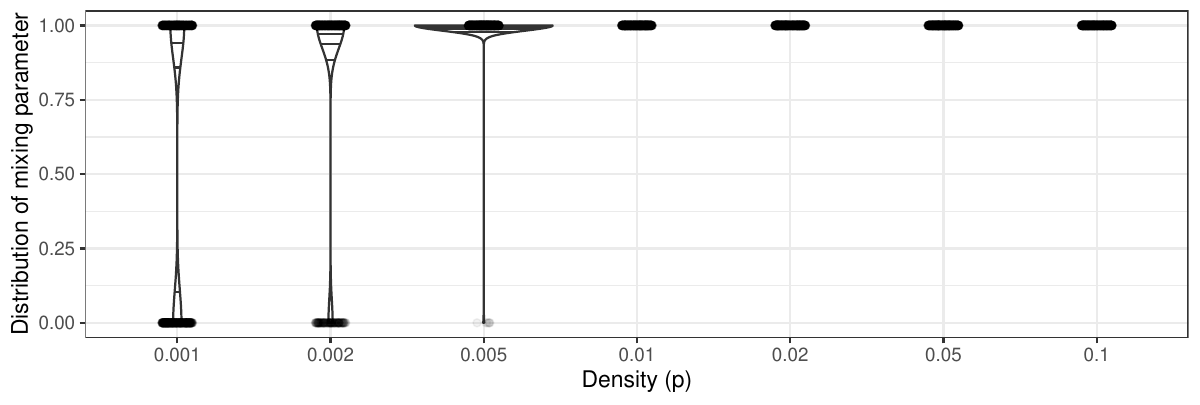}\\
    B) Erd\H{o}s-R\'eyni+LFR\hfill~\\\vspace{-0pt}
\includegraphics[width=0.9\textwidth]{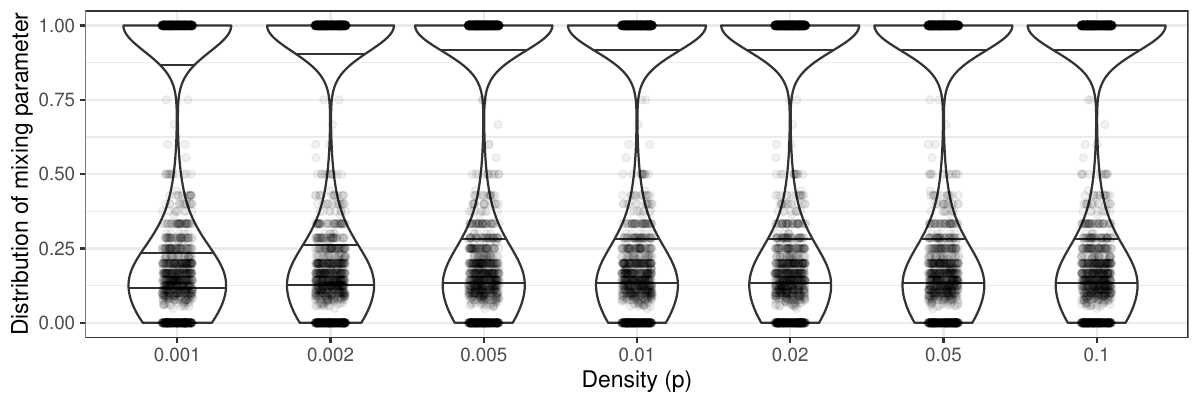}\\
    C) Erd\H{o}s-R\'eyni+Ring-of-cliques\hfill~\\\vspace{-0pt}
\includegraphics[width=0.9\textwidth]{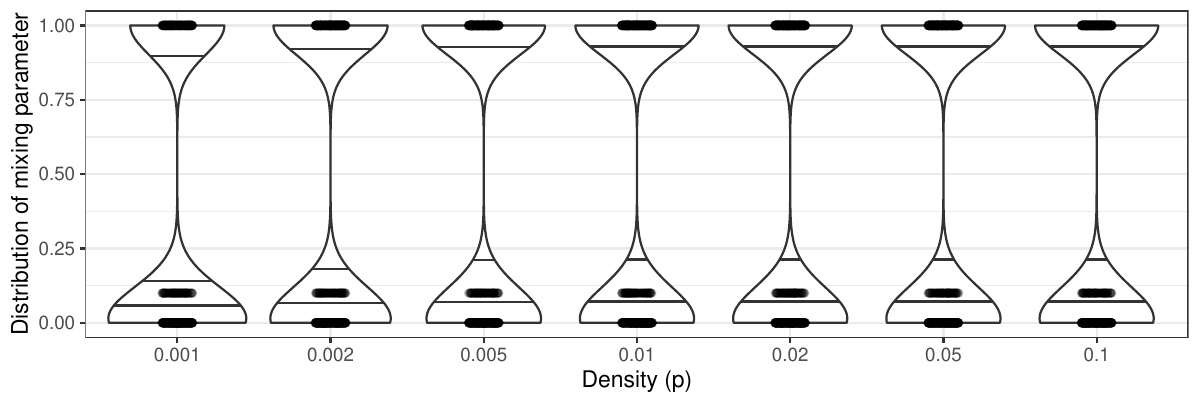}
    \caption[Distribution of the per-node mixing parameters for the networks used in Experiment 5. 
    ]{\textbf{Distribution of the per-node mixing parameters for networks used in Experiment 5.
    } A) Erd\H{o}s-R\'enyi graphs with size 1000 nodes and various densities.
    In figures B and C, the network is produced by combining an Erd\H{o}s-R\'eyni network with 1000 nodes to another network of 1000 nodes with very low mixing parameter. B) Distribution for Erd\H{o}s-R\'enyi+LFR networks.
    C) Distribution for Erd\H{o}s-R\'enyi combined with a  ring-of-cliques network, where each clique is of size 10.
    The x-axis specifies the density of the Erd\H{o}s-R\'enyi graph. 
    Mixing parameters are calculated with respect to the ground-truth community structure (no clusters of size greater than 1 in the Erd\H{o}s-R\'eyni graphs). 
     The 25\%, 50\% and 75\% quantiles are specified with black lines on the violin plots. The black points show the actual values of mixing parameters per-node that is variable for Erd\H{o}s-R\'enyi+LFR graphs, and take one of the three values of 0, 0.1 or 1 for Erd\H{o}s-R\'enyi+Ring-of-cliques (see Sec. B).
    }
\label{fig:mu-dist-er-lfr-ring}
\end{figure}

\begin{figure}[h!]
    \centering
    \includegraphics[width=1\textwidth]{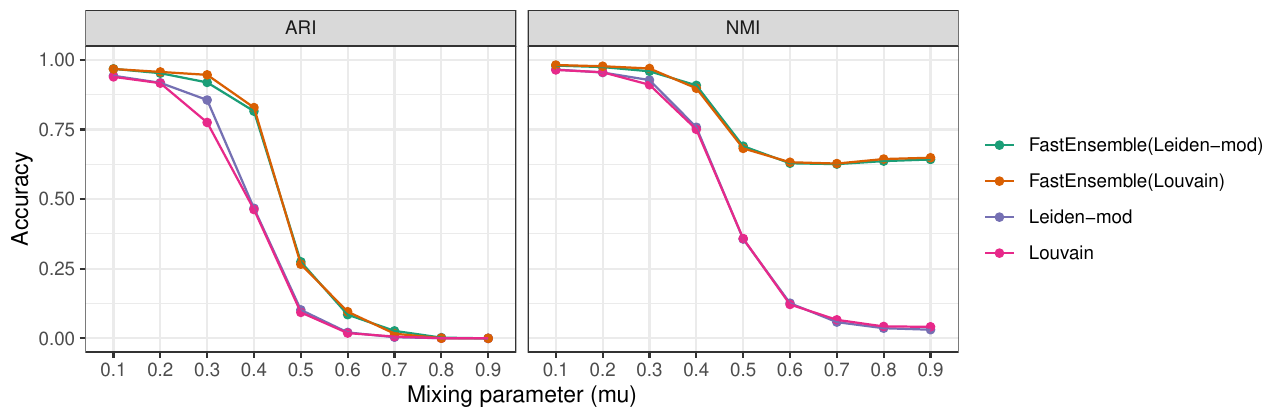}
    \caption[Comparison of Louvain and Leiden-mod on the algorithm design dataset  (Experiment 1b)]{\textbf{Comparison of Louvain and Leiden-mod on Algorithm Design datasets.} 
    The methods compared are Leiden, Louvain, and FastEnsemble used with these two methods. 
    The plots show accuracy of clustering methods on the algorithm design dataset from Experiment 1 for conditions with the default average degree (10) and network size (10,000); mixing parameters on the x-axis are the parameter values for generating the synthetic networks. 
    Note that Leiden-mod and Louvain have nearly identical accuracy under all conditions, FastEnsemble used with Leiden-mod is nearly identical to FastEnsemble used with Louvain, and FastEnsemble used with Louvain or Leiden-mod are more accurate than Louvain or Leiden-mod. 
    }
    \label{fig:expt-louvain}
\end{figure}

\begin{figure}[h!]
    \centering
    \includegraphics[width=1\textwidth]{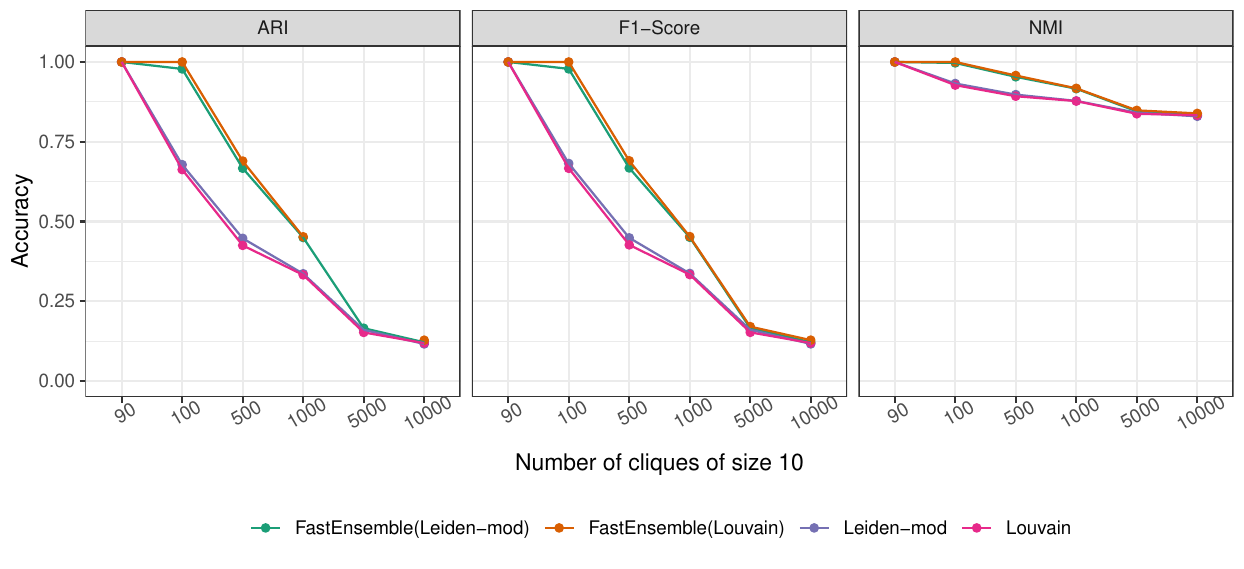}
     \caption[Comparison of Louvain and Leiden-mod  on Ring-of-Cliques networks (Experiment 4).]{\textbf{Comparison of Louvain  and Leiden-mod  on Ring-of-Cliques networks of various sizes}.  The methods compared are Louvain and Leiden-mod, as well as FastEnsemble used with these methods. 
     Each network consists of a varying number of cliques of size 10 arranged in a ring, where the total number of cliques ranges from 90 to 10,000 (specified on the x-axis). 
     Accuracy is shown using NMI, ARI and F1-score. Note that Leiden-mod and Louvain have nearly identical accuracy under all conditions, FastEnsemble used with Leiden-mod is nearly identical to FastEnsemble used with Louvain, and FastEnsemble used with Louvain or Leiden-mod is at least as accurate as Louvain or Leiden-mod. 
     }
    \label{fig:res_limit_exps_louvain}
\end{figure}


\begin{figure}[ht!]
    \centering
    \includegraphics[width=1\textwidth]{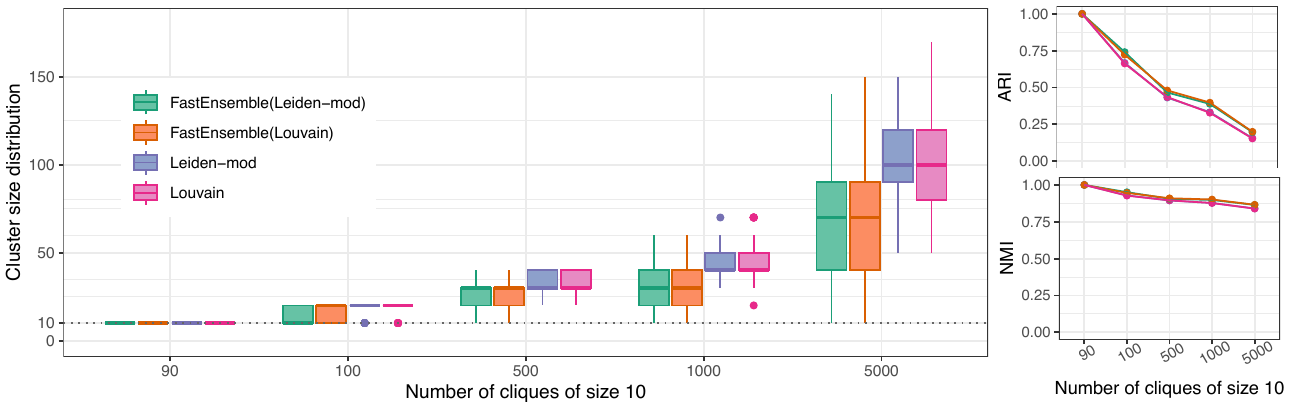}
    \caption[Comparison of Louvain and Leiden-mod on Tree-of-Cliques networks (Experiment 4).]{\textbf{Comparison of Louvain and Leiden-mod on Tree-of-Cliques networks.} 
    The methods compared are Louvain and Leiden-mod, as well as FastEnsemble used with each of these. 
    Each network consists of a varying number of cliques of size 10 arranged according to the structure of a random tree, where the total number of cliques ranges from 90 to 5,000 (specified on the x-axis). Leiden-mod and Louvain have nearly identical accuracy under all conditions, FastEnsemble used with Leiden-mod is nearly identical to FastEnsemble used with Louvain, and FastEnsemble used with Louvain or Leiden-mod is at least as accurate as Louvain or Leiden-mod. 
    }
\label{fig:res_limit_exps_lieden_mod_tree}
\end{figure}

\clearpage
\bibliographystyle{apalike}
\bibliography{ref}